\documentclass[longauth]{aa}  

\usepackage{graphicx}
\usepackage{natbib}
\bibpunct{(}{)}{;}{a}{}{,} 
\usepackage{hyperref}
\hypersetup{colorlinks=true, citecolor=blue, linkcolor=blue}
\usepackage{txfonts}
\newcommand{\um}{\mbox{$\mathrm{\mu m}$}}
\newcommand{\degree}{\mbox{$^{\circ}$}}
\newcommand{\roi}{\mbox{$R_\mathrm{OI}$}}
\newcommand{\trm}{\mbox{$\tau_\mathrm{RM}$}}
\newcommand{\rrm}{\mbox{$R_\mathrm{RM}$}}
\newcommand{\tsub}{\mbox{$T_\mathrm{sub}$}}
\newcommand{\rsub}{\mbox{$r_\mathrm{sub}$}}
\newcommand{\ledd}{\mbox{$\lambda_\mathrm{Edd}$}}
\newcommand{\rl}{\mbox{$R$--$L$}}
\newcommand{\rlbol}{\mbox{$R$--$L_{\mathrm{bol}}$}}
\newcommand{\lopt}{\mbox{$\lambda L_\lambda(\mathrm{5100 \AA})$}}
\newcommand{\lmir}{\mbox{$\lambda L_\lambda(12 \mathrm{\mu m})$}}
\newcommand{\lxray}{\mbox{$L_{14-195 \mathrm{keV}}$}}
\newcommand{\lbol}{\mbox{$L_{\mathrm{bol}}$}}
\newcommand*\samethanks[1][\value{footnote}]{\footnotemark[#1]}
\begin{document}

\title{VLTI/GRAVITY Interferometric Measurements of Innermost Dust Structure Sizes around AGNs}

   \author{
   GRAVITY Collaboration
\thanks{GRAVITY is developed in a collaboration by the Max Planck Institute for Extraterrestrial Physics, LESIA of Observatoire de Paris/Université PSL/CNRS/Sorbonne Université/Université de Paris and IPAG of Université Grenoble Alpes/CNRS, the Max Planck Institute for Astronomy, the University of Cologne, the CENTRA – Centro de Astrofisicae Gravitação, and the European Southern Observatory.}:
A.~Amorim\inst{1,2}
\and G.~Bourdarot\inst{3}
\and W.~Brandner\inst{4} 
\and Y.~Cao\inst{3}\thanks{Corresponding authors:Y.~Cao (ycao@mpe.mpg.de) and J.~Shangguan (shangguan@mpe.mpg.de)}
\and Y.~Cl\'enet\inst{5} 
\and R.~Davies\inst{3}
\and P.~T.~de~Zeeuw\inst{6} 
\and J.~Dexter\inst{3,7}
\and A.~Drescher\inst{3}  
\and A.~Eckart\inst{8,9} 
\and F.~Eisenhauer\inst{3} 
\and M.~Fabricius\inst{3}
\and H.~Feuchtgruber\inst{3}
\and N.~M.~F\"orster~Schreiber\inst{3} 
\and P.~J.~V.~Garcia\inst{2,10,11} 
\and R.~Genzel\inst{3,12} 
\and S.~Gillessen\inst{3} 
\and D.~Gratadour\inst{5,13} 
\and S.~H\"onig\inst{14}
\and M.~Kishimoto\inst{15} 
\and S.~Lacour\inst{5,16} 
\and D.~Lutz\inst{3} 
\and F.~Millour\inst{17}  
\and H.~Netzer\inst{18} 
\and T.~Ott\inst{3} 
\and K.~Perraut\inst{19} 
\and G.~Perrin\inst{5}
\and B.~M.~Peterson\inst{20}
\and P.~O.~Petrucci\inst{19} 
\and O.~Pfuhl\inst{16}
\and M.~A.~Prieto\inst{21}  
\and S.~Rabien\inst{3}
\and D.~Rouan\inst{5}
\and D.~J.~D.~Santos\inst{3}
\and J.~Shangguan\inst{3}\samethanks[2]
\and T.~Shimizu\inst{3}
\and A.~Sternberg\inst{18,22} 
\and C.~Straubmeier\inst{8} 
\and E.~Sturm\inst{3} 
\and L.~J.~Tacconi\inst{3} 
\and K.~R.~W.~Tristram\inst{10}  
\and F.~Widmann\inst{3} 
\and J.~Woillez\inst{16}
   }
\institute{
Universidade de Lisboa - Faculdade de Ci\^{e}ncias, Campo Grande, 
1749-016 Lisboa, Portugal
\and CENTRA - Centro de Astrof\'isica e Gravita\c{c}\~{a}o, IST, Universidade de Lisboa, 
1049-001 Lisboa, Portugal
\and Max Planck Institute for Extraterrestrial Physics (MPE), Giessenbachstr.1, 
85748 Garching, Germany
\and Max Planck Institute for Astronomy, K\"onigstuhl 17, 69117, Heidelberg, Germany
\and LESIA, Observatoire de Paris, Universit\'e PSL, CNRS, 
Sorbonne Universit\'e, Univ. Paris Diderot, Sorbonne Paris Cit\'e, 5 place Jules Janssen, 
92195 Meudon, France
\and Leiden University, 2311EZ Leiden, The Netherlands
\and Department of Astrophysical \& Planetary Sciences, JILA, University of Colorado, 
Duane Physics Bldg., 2000 Colorado Ave, Boulder, CO 80309, USA
\and I. Institute of Physics, University of Cologne, Z\"ulpicher Stra{\ss}e 77, 
50937 Cologne, Germany
\and Max Planck Institute for Radio Astronomy, Auf dem H\"ugel 69, 53121 Bonn, Germany
\and European Southern Observatory, Alonso de C\'ordova 3107, Casilla 19001, Vitacura, Santiago, Chile
\and Faculdade de Engenharia, Universidade do Porto, rua Dr. Roberto Frias, 
4200-465 Porto, Portugal
\and Departments of Physics and Astronomy, Le Conte Hall, University of California, 
Berkeley, CA 94720, USA
\and Research School of Astronomy and Astrophysics, Australian National University, 
Canberra, ACT 2611, Australia
\and Department of Physics and Astronomy, University of Southampton, Southampton, UK
\and Department of Physics, Kyoto Sangyo University, Kita-ku, Japan
\and European Southern Observatory, Karl-Schwarzschild-Str. 2, 85748 Garching, Germany
\and Universit\'e C\^ote d'Azur, Observatoire de la C\^ote d'Azur, CNRS, 
Laboratoire Lagrange, Nice, France
\and School of Physics and Astronomy, Tel Aviv University, Tel Aviv 69978, Israel
\and Univ. Grenoble Alpes, CNRS, IPAG, 38000 Grenoble, France
\and Retired
\and Instituto de Astrof\'isica de Canarias (IAC), E-38205 La Laguna, Tenerife, Spain
\and Center for Computational Astrophysics, Flatiron Institute, 162 5th Ave., 
New York, NY 10010, USA}

\titlerunning{Innermost dust structure sizes around AGNs}
\authorrunning{GRAVITY Collaboration}

\date{Received ; accepted }

 
  \abstract{  
We present new VLTI/GRAVITY near-infrared interferometric measurements of the angular size of the innermost hot dust continuum for 14 type 1 AGNs. The angular sizes are resolved on scales of $\sim$0.7~mas and the inferred ring radii range from 0.028 to 1.33 pc, comparable to those reported previously and a factor 10-20 smaller than the mid-infrared sizes in the literature. Combining our new data with previously published values, we compile a sample of 25 AGN with bolometric luminosity ranging from $10^{42}$ to $10^{47} \mathrm{erg~s^{-1}}$, with which we study the radius-luminosity (\rl) relation for the hot dust structure. Our interferometric measurements of radius are offset by a factor 2 from the equivalent relation derived through reverberation mapping. Using a simple model to explore the dust structure's geometry, we conclude that this offset can be explained if the 2~$\mu$m emitting surface has a concave shape. 
Our data show that the slope of the relation is in line with the canonical $R \propto L^{0.5}$ when using an appropriately non-linear correction for bolometric luminosity. In contrast, using optical luminosity or applying a constant bolometric correction to it  results in a significant deviation in the slope, suggesting a potential luminosity dependence on the spectral energy distribution.
Over four orders of magnitude in luminosity, the intrinsic scatter around the \rl~ relation is 0.2 dex, suggesting a tight correlation between innermost hot dust structure size and the AGN luminosity. }

\maketitle
%

\section{Introduction}
\label{sec:intro}

The central engine of an active galactic nucleus (AGN) is powered by accretion onto 
a supermassive black hole (SMBH) with a mass that can be in the range from $\sim$10$^5$~M$_\odot$ in dwarf galaxies \citep{Baldassare2015,Reines2022,Mezcua2024} to $\sim$10$^{10}$~M$_\odot$ in the most massive galaxies \citep{McConnell2011,Mehrgan2019}.
The UV photons from the accretion disk \citep{Shakura1973} ionize the gas in its close proximity, forming the so-called broad line region (BLR).  
Dust sublimates on these scales \citep{Barvainis1987}, but is an important component further out because of the key observational impact it has.
It is found in disk, outflow, and filament structures 
in the innermost region surrounding the AGN which is responsible for significant nuclear obscuration \citep{Antonucci1985,Urry1995,Honig2019,Prieto2021}.
Much of the progress in our understanding of these inner structures of AGN has come about through substantial improvements in observational techniques \citep{Netzer2015}. 
Notably, mid-infrared (MIR) and near-infrared (NIR) interferometry -- here referred to as optical/infrared interferometry (OI) -- enables one to spatially resolve sub-parsec scales even in distant objects and is opening new opportunities for studies of AGN.

Long-baseline infrared interferometry has made it possible to delve into 
the detailed structure of dust by resolving its thermal emission at different wavelengths.
At MIR wavelengths, interferometric observations of AGNs resolved warm dust structures emitting at a typical temperature of $400 \rm ~K$ on scales of 3--30~mas \citep{Kishimoto2011b, Burtscher2013}. 
Contrary to the classical torus model, detailed modelling of the data has revealed a significant fraction of the total flux coming, in many sources, from the polar region containing graphite grain dust on parsec or larger scales \citep{Honig2013,LopezGonzaga2016,Leftley2018}.

An alternate approach for investigating the dust structure is to monitor the time delay between the optical and 
NIR continuum emission. Using the time delay as an indicator of size, this reverberation mapping (RM) technique has measured the sizes of hot dust structures for $\sim 30$ AGNs 
\citep[e.g.][]{Clavel1989,Suganuma2006,Koshida2014,Minezaki2019}. 
At MIR wavelengths, the multi-epoch measurements of the WISE satellite 
\citep{Wright2010} are also an effective way to measure time lags \citep{Lyu2019,Yang2020, Mandal2024}. 
These efforts have not only confirmed the general picture that the hot dust is outside the BLR \citep{Clavel1989,Gravity2023}, but also that the time lag is smaller than predicted for the sublimation 
radius of standard ISM dust composition and grain sizes. This offset implies the presence 
of large graphite dust grains and/or anisotropic illumination in the innermost region of the dusty structure \citep{Kishimoto2007}.  
In addition, the relation between the hot dust radius and AGN luminosity is found to be roughly $R \propto L^{0.5}$ \citep{Suganuma2006,Kishimoto2007}, as expected for the simplest theoretical scenarios. 
Recent work indicates that the \rl\ relation may be 
slightly shallower than the power of 0.5, although the physical reason for such a deviation is under debate \citep{Minezaki2019,SobrinoFigaredo2020}.

The sub-parsec NIR emission of typical 
type~1 AGNs has been successfully resolved with the Keck Interferometer \citep{Swain2003,Kishimoto2009,Pott2010} and the Very Large Telescope 
Interferometer (VLTI; \citealt{Weigelt2012}).
The sub-milliarcsec scales can be resolved by measuring the descrease of the visibility toward larger $uv$ distance -- which can be achieved with the baselines of 85~m for the Keck telescopes and 47--130~m for the Unit Telescopes (UTs) of the VLTI.
More recently, 
\cite{Kishimoto2022} reported a new measurement of NGC~4151 at even longer baselines of 
$\sim$250~m using the CHARA array.
The second-generation VLTI instrument GRAVITY \citep{Gravity2017}, simultaneously combining the light from the four 8-m UTs over their size baselines, has vastly improved sensitivity and $uv$ coverage.
It has been able to spatially resolve the BLR of AGNs for the first time not only at low redshift \citep{Gravity2018,Gravity2020,Gravity2021a,Gravity2021b,Gravity2024} 
but also, with the recent upgrades toward 
GRAVITY+ \citep{Gravity+2022}, at $z\sim2$ \citep{Gravity+2024}.  
In terms of hot dust emission, the first resolved image of the type~2 Seyfert NGC~1068 showed that it originates in a disk \citep{Gravity2020}, and combining this with mid-infrared interferometric data reveals it to be part of a disk plus outflow system \citep{GamezRosas2022,Leftley2023}.

NIR dust structure size measurements for about 10 AGNs, comparable to the total number observed previously, have been recently reported \citep{Dexter2020,Gravity2020,Leftley2021}.
Combined analyses of the available data have confirmed that the interferometric dust size ($\roi$) also follows a relation close to $R \propto L^{0.5}$, but that it is a factor 2 larger than the size inferred from the RM time lag (\trm)\footnote{Hereafter, we use the term time lag exchangeably with $\rrm \equiv c\trm$ for simplicity.}\citep{Kishimoto2011,Koshida2014,Dexter2020,Gravity2023}.
The difference is not unexpected because the OI size reflects the projected light distribution, while the RM size includes additionally the response of the hot dust emission to the central heating source \citep{SobrinoFigaredo2020}. 
\textcolor{black}{Moreover, RM is biased toward more compact structures as they respond more coherently than extended structures.}

Here we take this work a step further by reporting new GRAVITY measurements of hot dust continuum sizes of 14 low-$z$ AGNs, significantly enlarging the interferometric AGN sample.  
We observed most of the targets with short exposures and reduced the data in a way that 
optimizes the continuum visibility (Section~\ref{sec:obs}).
We also carefully quantify 
the error budget on the measured size to take into account variations within a single 
night and between multiple nights (Section~\ref{sec:size}).  
The \rl\ relation based on the full sample of OI 
measurements is discussed in Section~\ref{sec:rl}.  
Employing a Monte Carlo model adaptable for exploring variations in observed OI and RM sizes with different dust structure geometries, we find a  bowl-shape emitting hot dust surface can quantitatively 
explain the observed difference between OI and  RM sizes. 
The geometric covering factor and NIR colors of our favored model are also consistent with the observations (Section~\ref{sec:model}).  Our main results are summarized in 
Section~\ref{sec:sum}.  We adopted the \cite{Planck2016} cosmology: 
$\Omega_m=0.308$, $\Omega_\Lambda=0.692$, and 
$H_0=67.8~\mathrm{km\,s^{-1}\,Mpc^{-1}}$.

%


\section{Observations and Data Reduction}
\label{sec:obs}

\begin{table*}
\caption{Observation Log.}
\label{tab:obs}
\begin{center}
\begin{tabular}{lccccccc}
\hline\hline
Name            & Date       & Observation  & Seeing      & Coherence & Strehl     &  $K$  & $V$   \\
                &            & Mode         & (")         & time (ms) &            & (mag) & (mag) \\
 (1)            & (2)        & (3)          & (4)         & (5)       & (6)        &  (7)  & (8)   \\ \hline
Akn~120         & 2019-11-07 & ON           & 0.37--0.76 &  5.4--10.2 & 0.04--0.09 &  10.8 & 13.9  \\
IC~4329A        & 2021-02-01 & ON           & 0.36--0.99 &   4.1--7.2 & 0.06--0.48 &  10.8 & 13.7  \\
                & 2021-03-01 & ON           & 0.43--0.78 &   3.7--9.4 & 0.04--0.13 &  10.6 &       \\
                & 2021-03-02 & ON           & 0.62--0.95 &   4.7--7.0 & 0.03--0.11 &  10.3 &       \\
                & 2021-03-31 & ON           & 0.70--0.82 &   3.1--4.1 & 0.06--0.14 &  10.3 &       \\
Mrk~1239        & 2021-01-30 & ON           & 0.38--0.73 &   3.2--9.4 & 0.03--0.19 &   9.9 & 14.4  \\
                & 2021-02-01 & ON           & 0.32--0.44 &  5.0--15.5 & 0.14--0.31 &  10.2 &       \\	
                & 2021-03-01 & ON           & 0.66--1.05 &   3.1--7.8 & 0.02--0.09 &  10.1 &       \\
                & 2021-03-02 & ON           & 0.46--0.85 &   4.6--6.5 & 0.02--0.12 &   9.5 &       \\
                & 2021-03-31 & ON           & 0.48--0.94 &   2.5--5.5 & 0.02--0.14 &   9.5 &       \\
Mrk~509         & 2021-07-25 & ON           & 0.50--0.80 &   2.5--4.2 & 0.05--0.23 &  11.2 & 13.1  \\
                & 2021-07-26 & ON           & 0.58--0.80 &   1.9--3.4 & 0.03--0.15 &  11.0 &       \\
NGC~7603        & 2021-07-26 & OFF          & 0.58--0.71 &   2.3--3.3 & 0.02--0.09 &  10.8 & 14.0  \\
PDS~456         & 2021-07-27 & ON           & 0.54--0.82 &   2.9--4.6 & 0.03--0.13 &  11.0 & 14.0  \\
PGC~89171       & 2021-07-26 & OFF          & 0.61--0.73 &   2.4--3.2 & 0.05--0.14 &  11.1 & 14.4  \\
UGC~545         & 2021-07-25 & OFF          & 0.64--0.75 &   1.7--2.5 & 0.03--0.06 &  10.7 & 14.0  \\
NGC~3227        & 2022-05-19 & OFF          & 0.60--0.76 &  7.0--10.1 & 0.03--0.08 &  11.0 & 11.8  \\
HE~1029$-$1401  & 2022-05-19 & OFF          & 0.44--0.55 &  6.1--11.0 & 0.18--0.50 &  12.2 & 13.9  \\
NGC~4593        & 2022-05-19 & OFF          & 0.36--0.44 &  9.0--10.7 & 0.10--0.22 &  11.6 & 13.2  \\
NGC 7469        & 2023-06-01 & OFF          & 0.48--0.76 &   4.9--6.1 & 0.04--0.07 &  11.1 & 12.3  \\
                & 2023-06-03 & OFF          & 0.31--0.44 &  5.5--11.9 & 0.05--0.09 &  11.2 &       \\
IRAS~13349+2438 & 2023-06-06 & OFF          & 0.57--0.70 & 11.0--13.4 & 0.07--0.13 &  10.1 & 15.0  \\
UGC~11763       & 2023-06-06 & OFF          & 0.54--0.83 &   5.8--7.0 & 0.10--0.17 &  11.4 & 14.6  \\
\hline
\end{tabular}
\end{center}
{
\textbf{Notes.} 
Col. (1) Target name; 
Col. (2) Observation date;
Col. (3) Observation mode, ``ON'' for on-axis mode with light spitted in half into the science and fringe tracker and ``OFF'' for off-axis mode with all of the light used by the fringe tracker;
Col. (4)--(6) Averaged seeing, coherence time, and Strehl ratio during the observation.
Col. (7) Nuclear $K$-band magnitude of the target measured by GRAVITY fringe tracker. 
The typical uncertainty is 0.2~mag based on our multiple exposures during the same night.
Col. (8) Total $V$-band magnitude from Simbad.
}
\end{table*}

\subsection{Observations} 
\label{ssec:obs}

The observation reported in this work come from two 
projects\footnote{Observations were made using the ESO Telescopes at the La Silla 
Paranal Observatory, program IDs 1103.B-0626 and 0109.B-0270.}: (1) The GRAVITY 
AGN Large Programme (PI: Sturm) has the primary goal of spatially resolving 
the broad-line region of bright ($K<11$) Seyfert 1 galaxies; (2) Our project 
focusing on the AGN hot dust continuum (PI: Davies). Due to the different 
primary science goals, the target observations adopted different setups. 
The observations of Akn~120, IC~4329A, Mrk~1239, Mrk~509, and PDS~456 were 
mainly to resolve the BLR with the spectro-astrometry technique. Therefore, 
we observed these targets with the single-field on-axis mode with light of 
the targets split in half into the science and fringe tracker (FT) channels, 
respectively. These targets were observed in multiple nights. The fringe tracker 
takes quick exposures (300 Hz) with only 6~channels over the $K$ band to measure 
the coherence flux of the AGN within the coherence time of the atmosphere and is 
used for phase referencing the coherent integration of the science channel, 
where we adopted the MEDIUM ($R=500$) spectral resolution to measure the broad 
emission line.  
For the rest of the sources in this work, we only need FT data 
to resolve the hot dust continuum; therefore, our continuum-focused observations 
adopted the single-field off-axis mode with all of the target light used by 
the FT. Since the fringe tracker is limited by the brightness of the target, 
the single-field off-axis mode enables the observation of fainter targets than 
the single-field on-axis mode.  We can usually measure the continuum size of 
a target with the single-field off-axis mode in one epoch with a $\sim 1$ hour 
observation.  The observation information of our targets are summarized in 
Table~\ref{tab:obs}.



\subsection{Data Reduction}
\label{ssec:red}

We resolve the spatial extension of the hot dust continuum by measuring the drop 
of the visibility amplitude toward longer baselines.
We first reduce the raw data of the AGN and the calibrator using the Python 
tool, \texttt{run\_gravi\_reduce}, of the GRAVITY pipeline \citep{Lapeyrere2014} 
with all the default options except \texttt{----gravity\_vis.p2vmreduced-file=TRUE}.  
The latter option is used to generate the intermediate data products (i.e. 
the \texttt{P2VMRED} files) that consist of the uncalibrated visibility data of 
each short FT exposure and other auxiliary data, such as the group delay 
(\texttt{GDELAY}) and geometric flux (\texttt{F1F2}), that can be used to flag 
the low quality data.  

Previous works \citep{Dexter2020,Leftley2021} found coherence loss of 
the visibility correlated with the Strehl ratio during the observation.
This means that the measured visibility amplitude depends on the weather 
conditions and the performance of the adaptive optics.  \citet{Dexter2020} found 
that it is effective to select exposures with group delay $<3\,\um$ to 
alleviate the Strehl ratio dependence.  \citet{Leftley2021} found, however, it 
is more effective to select exposures with the highest 3\% geometric flux to 
alleviate the AO loss for the data of ESO~323-G77.  We tested both methods and find the selection based on the group delay performs better in 
general for our data. Using the group delay selection method, the rejection rate is substantially lower, typically around 50\%, peaking at a maximum of 85\%. Therefore, we choose to adopt the selection method of \citet{Dexter2020}.
We flag out the non-selected exposures in the \texttt{P2VMRED} files and use the 
Python tool \texttt{run\_gravi\_reduce\_from\_p2vmred} to generate the averaged 
uncalibrated visibility data.  We adopt the same data selection for both AGN and 
calibrator data, although it does not affect the calibrator data because most of 
the calibrator exposures have \texttt{GDELAY}$<3$~\um.  Finally, we use the Python tool 
\texttt{run\_gravi\_trend} to calibrate the AGN data with those of the calibrator, 
in order to remove the remaining instrumental effects of the visibility data.

%


\section{Hot dust size measurements}
\label{sec:size}

\begin{figure*}[htp]
\centering
\includegraphics[width=0.48\textwidth]{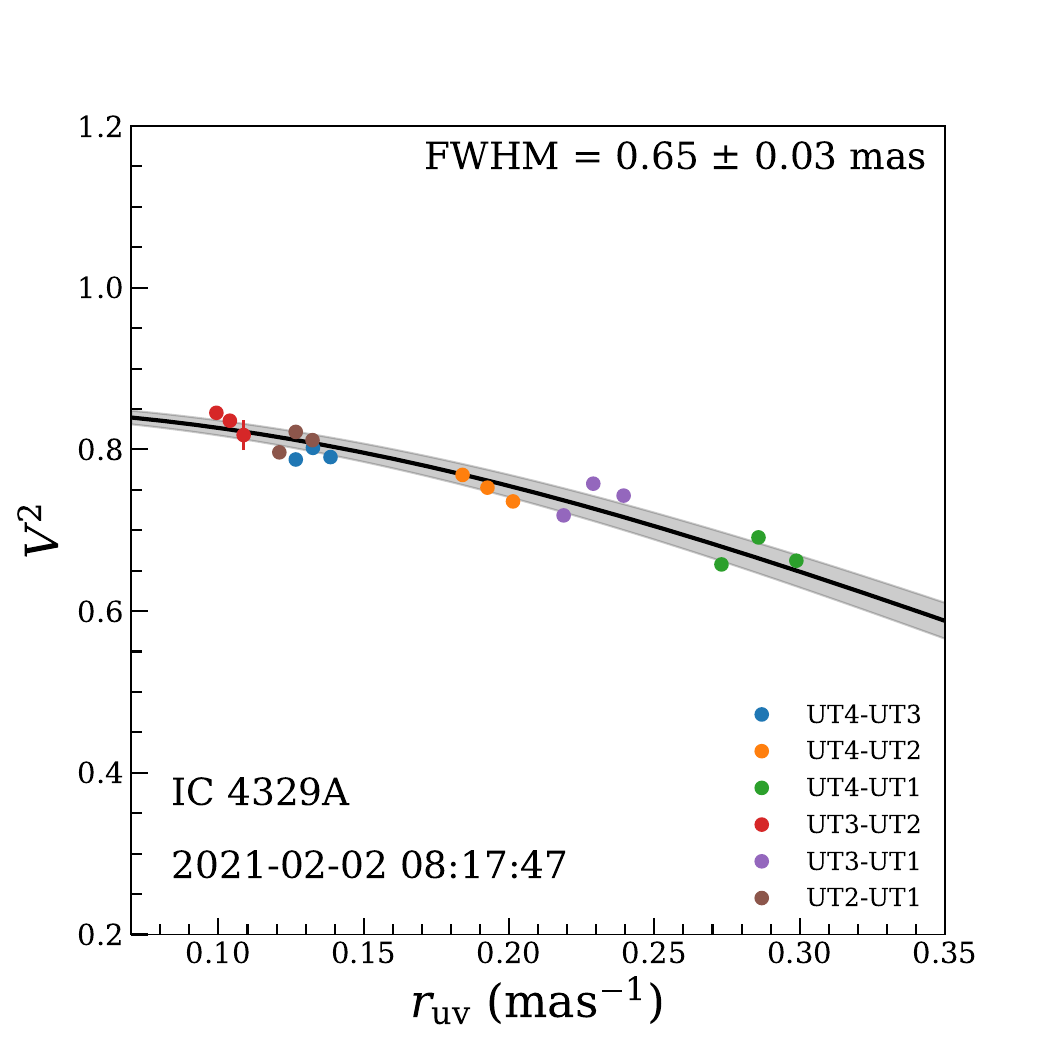}
\includegraphics[width=0.48\textwidth]{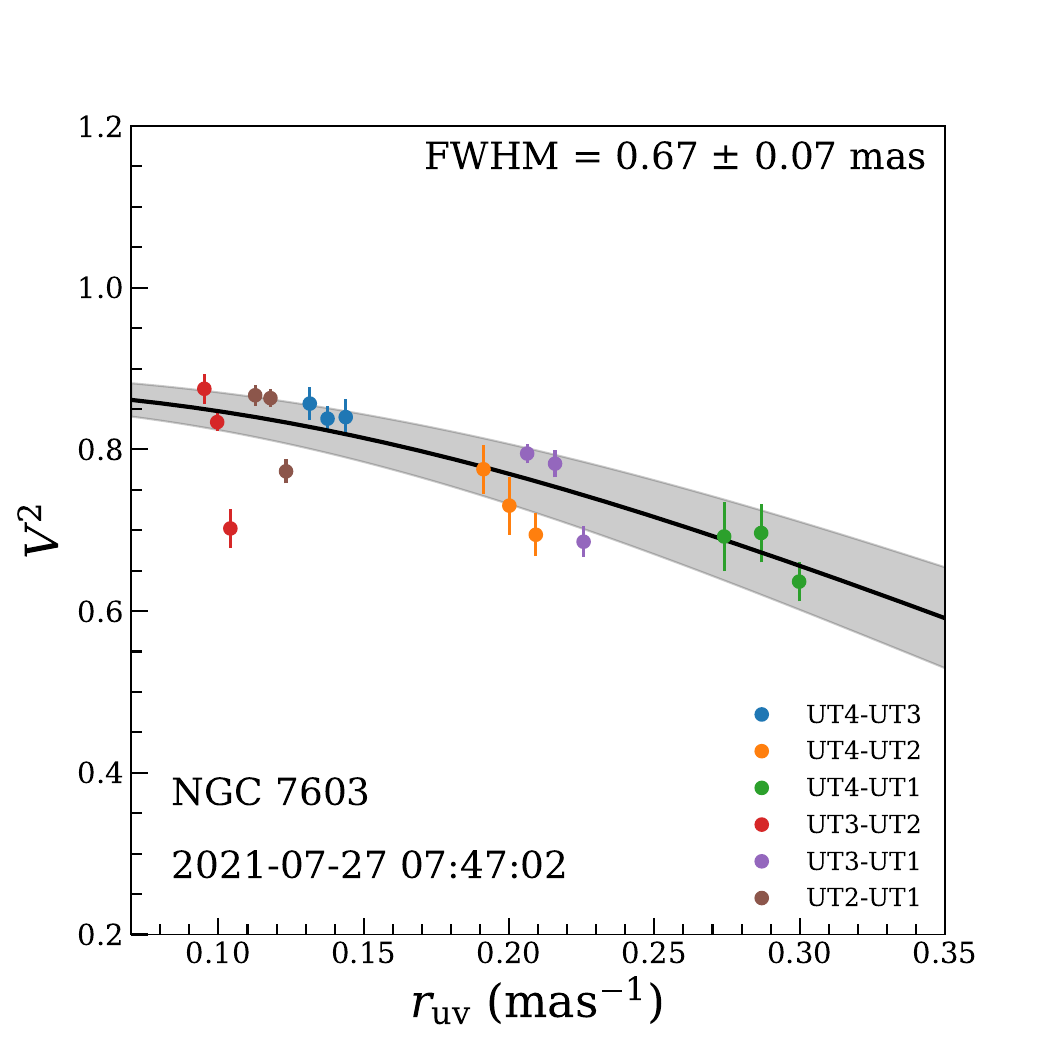}
\caption{FWHM visibility fitting for two single exposures. 
Colored dots with error bars show the visibilities measured at different baselines. Best fitting results from Equation \ref{eq:vsqare} are shown in black solid lines, with the gray shaded region show the 1-$\sigma$ fitting uncertainties. The dates and the UT time are shown in the left bottom corners of each panel.
{The left panel illustrates a typical exposure, while the right panel shows an example of poor quality data.}
}
\label{fig:vis-fit}
\end{figure*}

Interferometric resolution is determined by the baseline length $B$ between telescopes and the observed wavelength $\lambda$ as $\lambda/2B$ \citep{Eisenhauer2023}; in the case of the UTs of the VLTI, the resolution reaches $\sim$2~mas in K-band. 
However, one can measure the size of an object with about 10 times better resolution by \textcolor{black}{measuring} how the contrast (or visibility) of the interferometric fringes decreases with larger baseline length \citep{Dexter2020}. In doing this, the squared visibility 
(\texttt{VIS2DATA}, hereafter $V^2$) \textcolor{black}{is used }instead of the visibility amplitude (\texttt{VISAMP}) in 
the reduced GRAVITY FT data, because the former shows less coherence loss 
\citep{Dexter2020,Leftley2021}.

We measure the size of the hot dust by fitting the $V^2$ of the FT continuum 
data to a Gaussian model,
\begin{equation}\label{eq:vsqare}
     V^2 = V_{0}^{2} \exp{\left(\frac{-\pi^2 r_{uv}^2 \mathrm{FWHM}^2}{2\ln2}\right)} \ , 
\end{equation}
where $V_{0}$ is the zero baseline visibility, $r_{uv}$ is the baseline length 
in units of $\rm mas^{-1}$, and FWHM in mas is the full width at the half 
maximum of the source emission.  Following \cite{Dexter2020}, we also allow 
$V_0$ to be free in the fitting to account for the remaining coherence loss of 
the calibrated data and/or extended flux that is resolved out by the interferometer.  
We fit the visibility data from each individual exposure 
{(with $\sim 5$ minutes exposure time for the on-axis mode, and 2 minutes for the off-axis mode)}, 
and find the best-fitting zero baseline $V_{0}$ and FWHM using 
the \texttt{scipy} function \texttt{curve\_fit}.  Figure \ref{fig:vis-fit} shows 
two such examples.  The clear drops of $V^2$ with increasing baseline lengths 
indicate that the hot dust continuum is resolved.  We only incorporate 
the three central channels (2.07, 2.17, and 2.27~\um) of the FT data in 
the fitting because the remaining three channels are more susceptible to 
the detector background due to the metrology laser at the shorter wavelength and 
the thermal background at the longer wavelength.

We present the measured FWHM of each target in Table \ref{tab:sizes}. We specifically include only those exposures for which the fitted error is less than one-third of the FWHM value.  Subsequently, we calculate the median of these selected exposures to obtain the measured FWHM for each target. 
To estimate the FWHM 
uncertainty, \cite{Dexter2020} use the RMS of the FWHM from individual 
exposures and divide it by the square root of the number of nights.  In this 
way, they account for the night-to-night systematic uncertainty, which likely 
comes from the variation of the AO performance. This method, however, cannot be 
applied to most of our targets because they were only observed once.  Using 
the AGNs observed in multiple epochs to investigate the night-to-night FWHM 
variation, we find that it is about 10\% of the averaged FWHM of each night.  Therefore, we estimate the FWHM uncertainty by 
summing in quadrature two components: (1) the statistical uncertainty which is 
the RMS of FWHM values divided by the square root of the number of exposures; 
(2) the systematic uncertainty which is 10\% of the median FWHM.  Our method 
provides consistent FWHM uncertainties with that adopted by \cite{Dexter2020} 
for the targets observed with multiple nights.  

Next, we convert the fitted Gaussian FWHM to a physical continuum radius.  
Following \cite{Dexter2020}, we first convert the Gaussian FWHM to a ring radius 
by dividing \textcolor{black}{by} a factor of $2\sqrt{\ln 2}\approx 1.67$.
{
We then correct for a putative 
contamination due to the unresolved central source (the accretion
disk and/or jet)  with a flux fraction of $f$, by scaling up the ring radius by a factor of $1/\sqrt{1-f}$.
The flux fraction $f$ differs for each object, and we use a typical constant value $f=20\%$ \citep{Kishimoto2009} in our conversions into physical radii. 
Variation in $f$ between different individual sources would introduce a small uncertainty ($<10\%$) to the derived sizes.} 
 The results are reported in 
Table~\ref{tab:sizes}.

At the end of the table, we provide updated measurements for two sources previously published in \citet{Dexter2020}. The updated FWHM values remain consistent with the previous measurements ($0.59 \pm 0.08$ for PDS 456 and $0.54 \pm 0.06$ for Mrk 509). The reduction in uncertainties is attributed to the acquisition of additional exposures in 2021. Specifically, the updated sizes are based on approximately 1.5 times the number of exposures used for the previous published measurements.

%
\begin{table}[htp]
\caption{Angular and physical size measurements.}
\label{tab:sizes}
\begin{center}
\begin{tabular}{l c c c r}
\hline
\hline
\noalign{\smallskip}
Source   & FWHM (mas) & r (pc) 
\\
\noalign{\smallskip}
\hline
\noalign{\smallskip}
Akn 120 & $0.70 \pm 0.09 $  & $0.328 \pm 0.040 $  \\
HE 1029-1401 & $0.64 \pm 0.07 $  & $0.739 \pm 0.081 $  \\
IC 4329A & $0.64 \pm 0.02 $  & $0.151 \pm 0.004 $  \\
IRAS 13349+2438 & $0.76 \pm 0.08 $  & $1.075 \pm 0.108 $  \\
Mrk 1239 & $0.55 \pm 0.04 $  & $0.159 \pm 0.012 $  \\
NGC 3227 & $0.50 \pm 0.06 $  & $0.028 \pm 0.004 $  \\
NGC 4593 & $0.37 \pm 0.04 $  & $0.045 \pm 0.005 $  \\
NGC 7469 & $0.74 \pm 0.04 $  & $0.178 \pm 0.008 $  \\
NGC 7603 & $0.67 \pm 0.09 $  & $0.279 \pm 0.039 $  \\
PGC 89171 & $0.65 \pm 0.07 $  & $0.254 \pm 0.027 $  \\
UGC 11763 & $0.75 \pm 0.08 $  & $0.652 \pm 0.066 $  \\
UGC 545 & $0.70 \pm 0.07 $  & $0.594 \pm 0.061 $  \\
\hline
PDS 456 & $0.60 \pm 0.03 $  & $1.330 \pm 0.073 $  \\
Mrk 509 & $0.62 \pm 0.03 $  & $0.304 \pm 0.017 $  \\
\hline 
\end{tabular}
\end{center}
{
\textbf{Notes.} Angular FWHM sizes measured in this work, and corresponding physical sizes converted using a ring model. PDS 456 and Mrk 509 were published in \citet{Dexter2020} and we update the measured sizes combining both the new and previous observations. 
} 
\end{table}
\section{Dust radius-luminosity relation}
\label{sec:rl}

\begin{figure}[htp]
       \centering
       \includegraphics[width=0.48\textwidth]{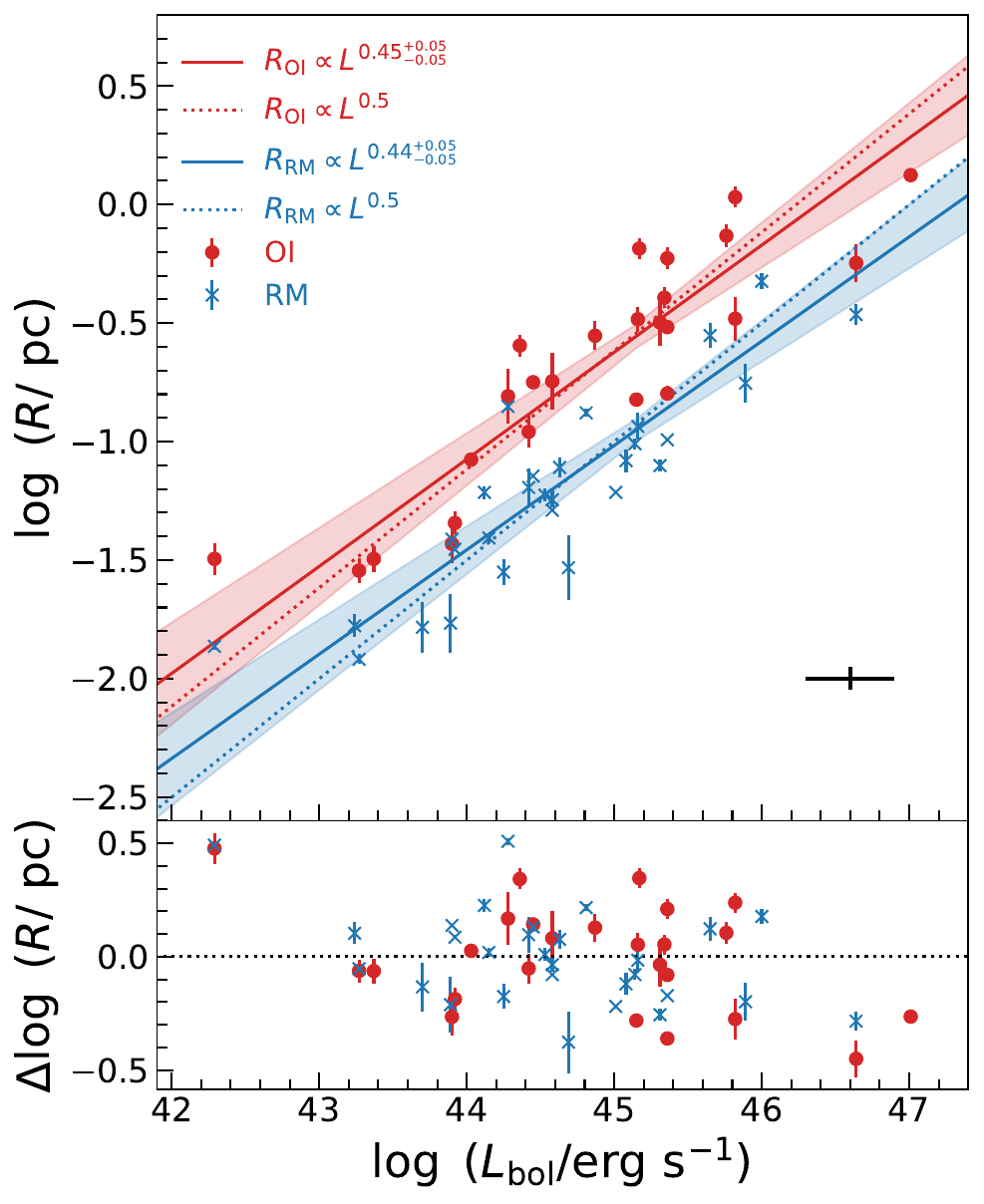}
       \caption{
       Dust radius as a function of bolometric luminosity (\rlbol~relation). OI measured sizes are shown in red and RM measured sizes are shown in blue. 
       {The typical 0.3 dex uncertainty in $L_{\rm bol}$ is indicated in the right corner. }
       {The solid lines show our best-fit results, with the shaded regions showing the 1-$\sigma$ uncertainties of the fittings, while the dotted lines represent the fitting results with the slopes $m$ fixed to 0.5,. Our best-fitted \rlbol~relations are consistent with the slope of 0.5 within 1-$\sigma$. 
       }
        {The bottom panel shows the dust radius residuals from the fitted relation with fixed slopes of 0.5.}
       }
       \label{fig:r-lbol}
\end{figure}

\begin{figure*}
\centering
\includegraphics[width=\textwidth]{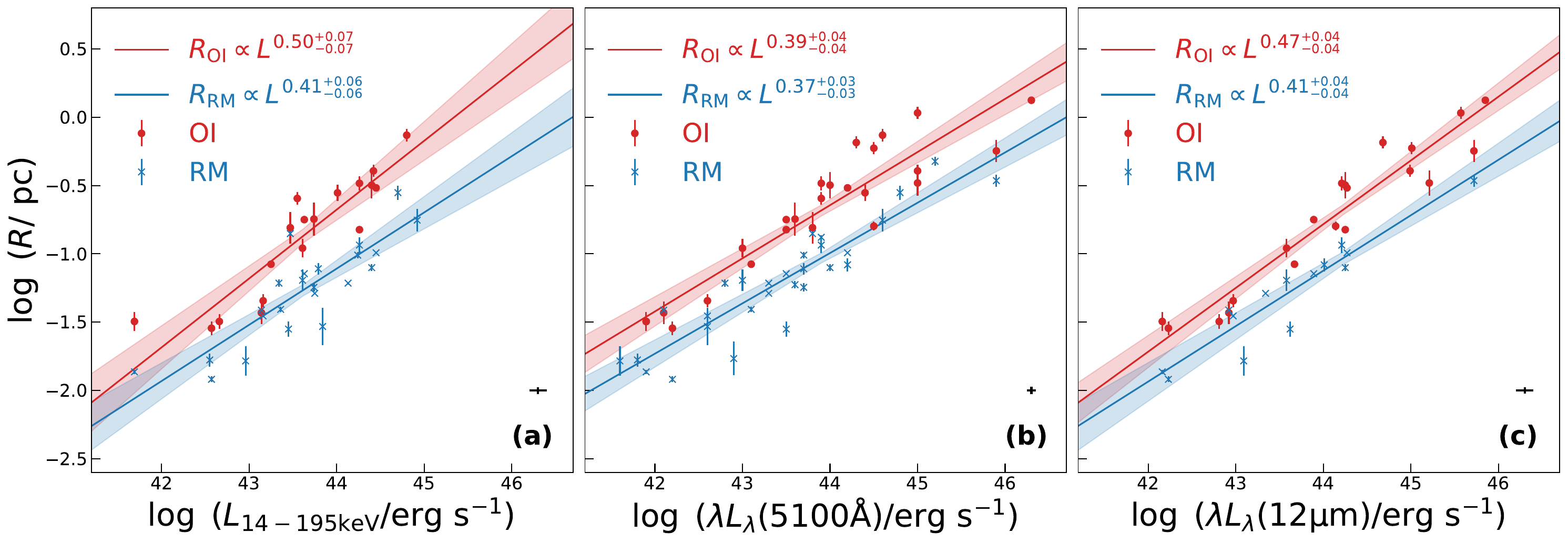}
\includegraphics[width=\textwidth]{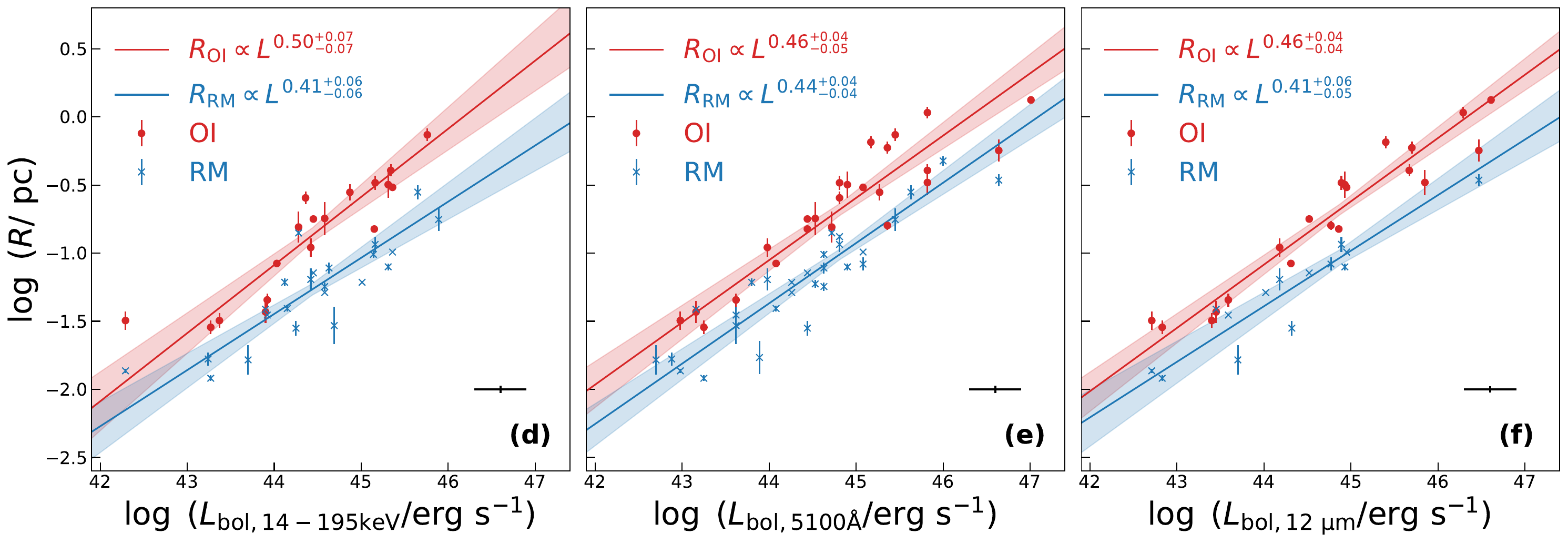}
\caption{ 
The top panels show the relations between the dust size and the monochromatic 
luminosities at (a) 14-195~keV, (b) optical 5100 $\AA$, and (c) 12~\um.  
The lower panels (d)--(f) show the \rl\ relation of the bolometric luminosity converted from these three monochromatic luminosities respectively.  The symbols are 
the same as Figure~\ref{fig:r-lbol}.  The slopes of the relations using RM and 
OI measurements agree with each other in all kinds of luminosities we investigated.  
For OI measurements, all the \rl~ relationships agree within one sigma with a slope of 0.5, except for the optical luminosity \lopt. 
}
\label{fig:r-l-all}
\end{figure*}




We measured the size of 12 new targets and updated 2 previous measurements with GRAVITY in this work. 
 In Table \ref{tb:lit-sizes} we compile from the the literature all other dust size OI measurements in the K band obtained so far. 
The last four targets are observed by Keck, while all the others are observed by GRAVITY. 
IRAS 13349+2438 has been observed by both Keck and GRAVITY; the ring size we measure with GRAVITY is consistent with  Keck measurement of $ 0.92 \pm 0.06~ \mathrm{pc}$  reported by \citet{Kishimoto2009}. 
We choose to adopt the more recent result measured by GRAVITY for this study. 
The compiled data set allows us to study the dust radius-luminosity (\rl) relation. 
Together with the literature results, we compiled a sample of 25 type~1 AGNs with
OI measured host dust structure sizes. For comparison, we also included 29 AGNs 
with an RM measured continuum size collected in Table~1 of our companion 
paper, \cite{Gravity2023}.  In that paper, we collected the latest 
measurements of the hot dust continuum by OI and RM for $z \lesssim 0.2$ AGNs, 
and our main focus was to investigate the relation between the BLR and the dust 
continuum size.  Throughout this work, we adopted the  
continuum size directly converted from the time delay without applying any redshift correction 
\citep[e.g.][]{Minezaki2019}, because the wavelength dependence of continuum 
emission size is expected to be less than $10\%$ for our low-z sample \citep{Gravity2023}.

We collected the AGN luminosity from the literature following the method 
introduced in \cite{Dexter2020}.  Briefly, we collect the AGN 14--195~keV X-ray 
(\lxray), optical (5100~\AA, \lopt), and 12~\um\ (\lmir) monochromatic 
luminosities whenever available.  The X-ray luminosity comes from the Swift/BAT 
observations \citep{Baumgartner2013}. 
We discarded the X-ray luminosities of 3C 273 and PDS 456 due to contamination from jet emission and significant variability, respectively.
The optical luminosities are taken from 
\cite{Gravity2023}, and the \lmir\ comes from the high resolution MIR 
observations by \cite{Asmus2011}.  

{
Fundamentally the dust is heated by the central optical/UV continuum source, and thus the size is expected to depend on the total optical-to-UV luminosity \citep{Barvainis1987}. However, measuring optical-to-UV luminosity is challenging due to various factors. We opt to use the bolometric luminosity that can be derived from various methods.}  
As detailed in Appendix~\ref{apd:lbol}, we calculated $L_{\rm bol}$ based on 14--195~keV measurements whenever possible, using non-linear corrections. 
In cases where such measurements are unavailable, we opt for values 
derived from the optical luminosity (see Table \ref{tab:lbol_use} for details).  
We include the \lmir\ to provide additional comparison to quantify 
the uncertainty of the bolometric luminosities and the \rl\ relations.
The differences in \lbol~estimates from different monochromatic 
luminosities agree to within 0.3~dex, which we adopt as the uncertainty.

We fit the \rl\ relations of the bolometric luminosity and various monochromatic 
luminosities in the form of Equation \ref{eq:rlfit} using the \texttt{linmix} 
package, which is a Python version of  \texttt{LINMIX\_ERR} from \cite{Kelly2007}, 
\begin{equation}
\label{eq:rlfit}
    \log R  = c + m \log \left(\frac{L}{L_0}\right),
\end{equation}
where $c$ and $m$ are the regression intercept and slope respectively, $R$ is the measured dust continuum size in pc, $L$ is the luminosity, and $L_0$ is fixed at the same value for both the OI and RM relations to reduce the degeneracy of the regression coefficients. 
In \texttt{LINMIX\_ERR}, the probability distribution of the independent variable is modeled as a mixture of $K$ Gaussian functions, and we choose $K = 2$ for all of our fittings throughout this work. 
We use the median values of the $c$ and $m$ coefficients obtained from the likelihood distributions generated by \texttt{LINMIX\_ERR} as the best-fit regression coefficients. The uncertainties associated with these coefficients are determined from the $16\%$ and $84\%$ percentiles of their likelihood distributions.
We also characterise the intrinsic scatter of the linear regression, $\sigma_{\rm intrin}$, using the median value from its distribution.  The best-fit parameters with uncertainties are listed in Table \ref{tab:fits}.

We first fit the dust size $R$ as a function of bolometric luminosity.  
Figure~\ref{fig:r-lbol} shows the relation of \rlbol\ to the continuum sizes 
measured by the OI and RM, respectively. The best-fitted slope for \roi\ is 
about 0.45, slightly shallower than, but consistent within $\sim 1\sigma$ of, 
the expected $R \propto L^{0.5}$.  The best-fit slope for \rrm\ is entirely 
consistent with that of \roi.  The uncertainty of the slope can be further 
reduced by expanding the sample with high and low luminosities.  The intrinsic 
scatter for the \roi\ relation is 0.2~dex, similar to that of 
the \rrm\ relation.   We do not find a significant correlation between the offsets from the fitted dust \rl~relation and the Eddington ratio with the OI measurements.

Our compilation of the literature 
measurements cannot guarantee that the continuum size and the AGN luminosities 
are measured close in time. 
\citet{Kishimoto2013} showed in NGC 4151 that any change in luminosity will not immediately affect the measured dust sublimation radius, only if the change persists over several years. 
A potential explanation for this reduced response to AGN variability proposed by \cite{Honig&Kishimoto2011}, is the  ``snowball'' model, in which clouds only gradually sublimate at the inner edge of the torus.
In this case the dust size may be relatively constant with time in a typical AGN, and the AGN variability itself may be the dominant source of intrinsic scatter (see more discussion in 
\citealt{Gravity2023}).

We also fit the \rl\ relations using different monochromatic luminosities 
(\lxray, \lopt, and \lmir) and their
corresponding bolometric luminosity following the equations in Appendix \ref{apd:lbol}.    
Using monochromatic luminosities separates SED effects which potentially influence dust structure size, while converting them to \lbol~with appropriate bolometric corrections aligns \rl~relations across wavelengths, mitigating differences related to the observed luminosity's wavelength. 
The results for all these \rl~relations 
are shown in Figure~\ref{fig:r-l-all} and Table \ref{tab:fits}. 

As in Figure \ref{fig:r-lbol}, we find that the slopes of the OI-measured \rl\ relations consistently agree with those of the RM-measured relations for each type of luminosity in Figure \ref{fig:r-l-all}. 
For monochromatic luminosities shown in the top row of Figure \ref{fig:r-l-all}, 
the slopes of the relations vary with wavelength, probably due to the SED change  as a function of luminosity. 
The slopes of the \lopt\ \rl\ relation in panel (b) shows the most significant deviation from the 0.5 power law, which is close to 3-$\sigma$.  This finding is consistent with previous studies 
{from} 
\citet{Minezaki2019} and 
\citet{SobrinoFigaredo2020}.
{Various explanations have been discussed by these authors to interpret the observed shallower slope, such as anisotropic illumination by the accretion disk,  
non-trivial composition and geometry of the dust structure 
delayed dust sublimation responses 
and nonlinear correlations between optical and UV luminosities
For detailed discussions and references of these interpretations, we direct the readers to \citet{Minezaki2019} and 
\citet{SobrinoFigaredo2020}. 
}


On the other hand, we find that \rl\ relations using bolometric luminosities derived from optical luminosities with nonlinear bolometric corrections show slopes consistent with the canonical $R \propto L^{0.5}$ relation {within 1-$\sigma$} (notably panel (e) in Figure \ref{fig:r-l-all}). 
Indeed, for OI measurements, the slopes of   the \rl~ relations fitted using bolometric luminosities are all in line with canonical value of 0.5. 
Therefore we suggest that comparing the  slope of the \rl\ relation observed using monochromatic luminosities (or a simple linear bolometric correction) to the canonical value of 0.5 may be misleading.   This is because the relation between monochromatic luminosities and dust heating is sensitive to the SED shape.
{The adopted non-linear bolometric correction in our approach, which results in steeper slopes than the monochromatic \rl~relations, may reflect the systemic dependency of SED shape on luminosities \citep[e.g.][]{Vignali2003, Netzer2019, Duras2020}}. 

Recent RM studies using WISE $W1$ 
(3.4 \um) and $W2$ (4.6 \um) data support our conclusion.  Several works  
\citep{Chen2023,Mandal2024} reporting their \rl\ relation slope shallower than 
0.5,  either used the \lopt\ or a $V$-band luminosity with a constant bolometric 
correction.  In contrast, \cite{Lyu2019} applied a nonlinear bolometric 
correction, albeit different from our approach, and reported a slope similar 
to our result.  
Similar differences in slopes between \lopt~and \lbol~are also evident in the BLR \rl~relations \citep{Gravity+2024}.
For studying the \rl~ relations, we recommend using the more accurate nonlinear  correction \citep[e.g.,][]{Netzer2019, Duras2020} rather than a linear approximation for the bolometric luminosities.  

In all the fits of \rl\ relations we have examined, the intrinsic scatters are predominantly below 0.2 dex. Typically, the intrinsic scatters from monochromatic luminosities are slightly larger than those from bolometric luminosities.  
This may be due to the larger uncertainty assigned to the bolometric luminosities, while the larger scatter in the monochromatic luminosities likely reflects the specific characteristics of individual sources' SEDs. 
Contamination of NIR continuum from compact jet emission in some sources \citep{Fernandez-Ontiveros2023} 
could also contribute to the scatter in the \rl\ relations. 
Additionally, the intrinsic scatter is influenced by the range of luminosities and the number of data points used for the fitting. For instance, the higher intrinsic scatter from \lxray~ is likely due to the small range of luminosity, because of the lack of high luminosity objects in this sample. Additionally, a smaller range of luminosity also introduces a larger uncertainty on the fitted slopes. Nevertheless, the relatively small intrinsic scatter from all the relations indicates that the \rl\ relations are tightly constrained.





\begin{table}
\begin{center} 
\caption{Literature physical size measurements. }
\label{tb:lit-sizes}
\begin{tabular}{l c c c r}
        \hline
        \hline
        \noalign{\smallskip}
        Source   & $R_{\rm OI}$ (pc) &  Ref
        \\
        \noalign{\smallskip}
        \hline 
        \noalign{\smallskip}
   3C 120 & $0.318 \pm 0.071 $  & $1 $  \\
    3C 273 & $0.567 \pm 0.106 $  & $1 $  \\
    ESO 323-G77 & $0.084 \pm 0.004 $  & $2 $  \\
    IRAS 09149-6206 & $0.405 \pm 0.041 $  & $1 $  \\
    Mrk 335 & $0.155 \pm 0.041 $  & $1 $  \\
    NGC 1365 & $0.032 \pm 0.004 $  & $1 $  \\
    NGC 3783 & $0.110 \pm 0.017 $  & $1 $  \\
    \hline
    Mrk 231 & $0.330 \pm 0.070 $  & $3 $  \\
    Mrk 6 & $0.180 \pm 0.050 $  & $4 $  \\
    NGC 4051 & $0.032 \pm 0.005 $  & $3 $  \\
    NGC 4151 & $0.037 \pm 0.007 $  & $4 $  \\

    
        \hline
\end{tabular}
\end{center}
\textbf{References:}{
(1) \cite{Dexter2020}
(2) \cite{Leftley2021}
(3) \cite{Kishimoto2009}
(4) \cite{Kishimoto2011}
}
\end{table}

\begin{table*}[htb]
\caption{Results of linear regression of R-L relations for OI and RM measurements}
\label{tab:fits}
\begin{center}
\begin{tabular}{c c c c c c c c r}
\hline
\hline
\noalign{\smallskip}
 $L$ & Definition 
  & $\sigma_{\log L}$ 
& $\log(L_0/\rm erg \ s^{-1})$ 

 & $R$   &
$N$ & $c$ 
& $m$ & $\sigma_{\rm intrin}$
\\
\noalign{\smallskip}
\hline
\noalign{\smallskip}
  $L_{\rm bol}$& Bolometric luminosity in this work 
& 0.3 
& $45.1$  & OI
& $25$ & $-0.55_{-0.05}^{+0.05}$ & $0.45_{-0.05}^{+0.05}$ & $0.20$ \\

&  & &  & RM 
& $29$ & $-0.95_{-0.05}^{+0.05}$ & $0.44_{-0.05}^{+0.05}$ & $0.15$ \\

\noalign{\smallskip}
\hline
\noalign{\smallskip}

 $L_{\rm 14-195 \ keV}$&  $14-195 \rm \ keV$ luminosity  
 & 0.1 
& $43.6$  & OI 
 & $18$ & $-0.87_{-0.05}^{+0.05}$ & $0.50_{-0.07}^{+0.07}$ & $0.21$ \\

&  & &  & RM   
& $23$ & $-1.26_{-0.04}^{+0.04}$ & $0.41_{-0.06}^{+0.06}$ & $0.19$ \\

$\lambda L_{\lambda}(5100 \rm \ \AA)$ & Monochromatic luminosity at $5100 \rm \  \AA$ 
 & 0.05 
 & $43.9$ & OI 
 & $25$ & $-0.68_{-0.04}^{+0.04}$ & $0.39_{-0.04}^{+0.04}$ & $0.20$ \\
  &  &  & & RM 
& $29$ & $-1.03_{-0.04}^{+0.04}$ & $0.37_{-0.03}^{+0.03}$ & $0.17$ \\

  $\lambda L_{\lambda}(12 \rm \ \mu m)$ & Monochromatic luminosity at $12 \rm \ \mu m$  
   & 0.1 
& $44.2$ & OI
& $20$ & $-0.68_{-0.04}^{+0.04}$ & $0.47_{-0.04}^{+0.04}$ & $0.16$ \\
&  & &  & RM 
& $14$ & $-1.03_{-0.05}^{+0.05}$ & $0.41_{-0.04}^{+0.04}$ & $0.14$ \\

\noalign{\smallskip}
\hline
\noalign{\smallskip}
$L_{\rm bol, 14-195 \ keV}$ & 
Bolometric luminosity from $L_{\rm 14-195 \ keV}$ 
 & 0.3 
& $44.4$  &   OI
& $18$ & $-0.87_{-0.05}^{+0.05}$ & $0.50_{-0.07}^{+0.07}$ & $0.16$ \\
&   &  &   & RM
& $23$ & $-1.27_{-0.04}^{+0.04}$ & $0.41_{-0.06}^{+0.06}$ & $0.16$ \\

$ L_{\rm bol, 5100\AA}$ & 
Bolometric luminosity from $L_{\lambda}(5100 \rm \ \AA)$ 
 & 0.3 
& $44.8$ &   OI 
& $25$ & $-0.68_{-0.04}^{+0.04}$ & $0.46_{-0.05}^{+0.04}$ & $0.16$ \\
&   &   &  & RM
& $29$ & $-1.01_{-0.04}^{+0.04}$ & $0.44_{-0.04}^{+0.04}$ & $0.12$ \\

$L_{\rm bol, 12\mu m}$ &
Bolometric luminosity from $L_{\lambda}(12 \rm \ \mu m)$ 
 & 0.3 
& $44.9$  & OI
& $20$ & $-0.68_{-0.04}^{+0.04}$ & $0.46_{-0.04}^{+0.04}$ & $0.11$ \\
&   &  &  & RM
& $14$ & $-1.03_{-0.06}^{+0.05}$ & $0.41_{-0.05}^{+0.06}$ & $0.12$ \\

\hline
\end{tabular}
\end{center}
\textbf{Notes.}{ $R$ is the type of the dust size measurements. $N$ is the number of the data pair of each fitting, depending on the data availability of each kind of luminosity measurement. $L_{0}$, $c$, and $m$ are defined in Equation \ref{eq:rlfit}, and $\sigma_{\rm intrin}$ is the intrinsic scatter about the relation.}  
\end{table*}

\section{Constraining the hot dust structure}
\label{sec:model}

\begin{figure*}[htb]
\centering
\includegraphics[width=0.75\textwidth]{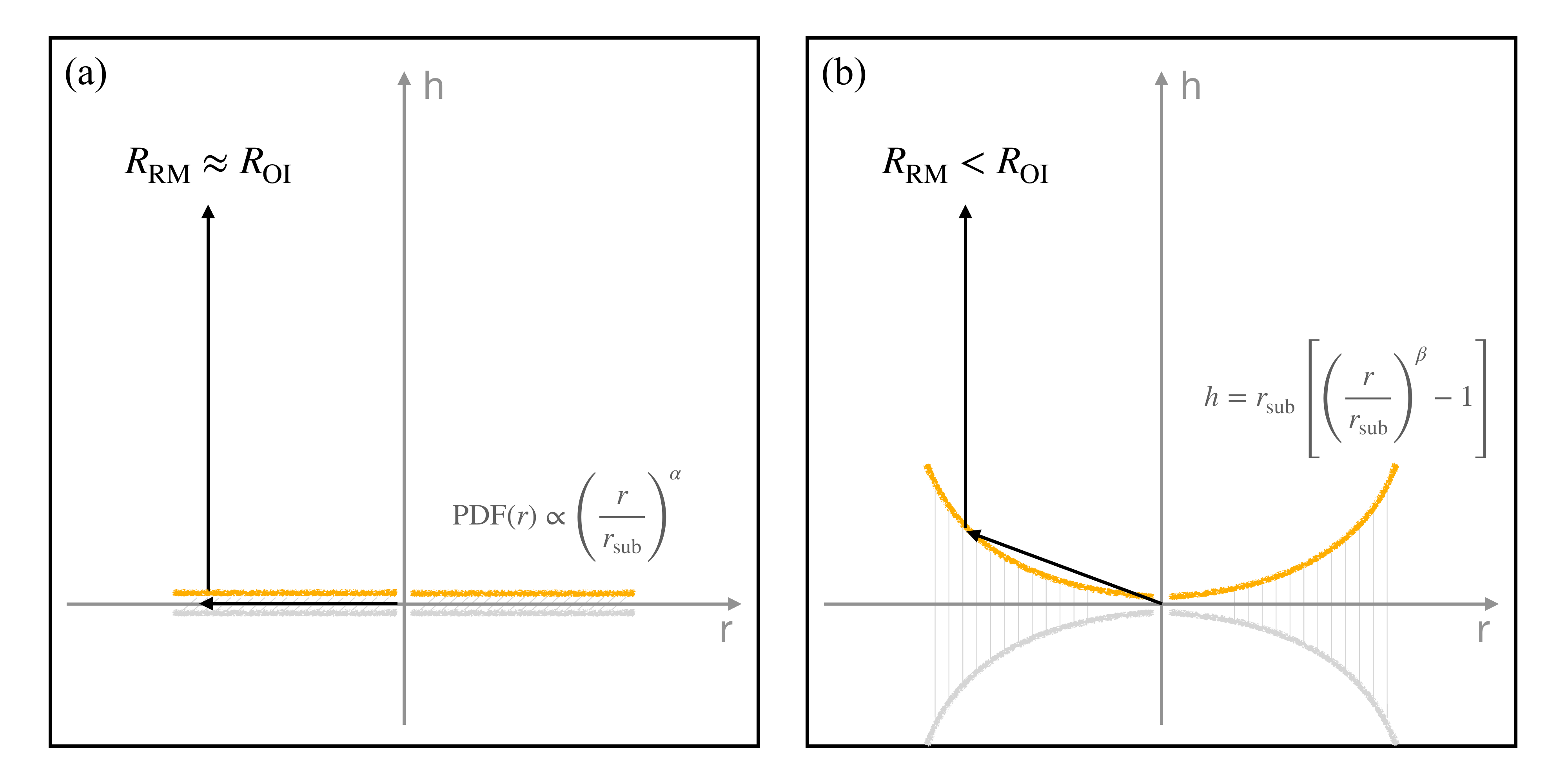}
\caption{The sketch of the hot dust structure model. The hot dust emission is in yellow.  
The cloud radial distribution follows a power law with the index $\alpha$, and 
the cloud height above the midplane is close to a power-law function controlled 
by $\beta$.  Panel~(a) shows the model with $\beta \approx 0$, while panel~(b) shows a model with $\beta>0$.  We show the models in an edge-on 
view with the observer to the positive direction of $h$ to illustrate 
the foreshortening effect. In this way, model (a) has $\rrm \approx \roi$, while 
model (b) has $\rrm < \roi$.}
\label{fig:sketch}
\end{figure*}

The \rl\ relations from both OI and RM measurements have similar slopes.  
However, there is a general offset between the two relations.  As shown in 
Table~\ref{tab:fits}, the intercepts ($m$) fitted from OI data are 
$\sim 0.3 \mathrm{-} 0.4 \rm \ dex$ larger than those from the RM method.  This means 
the hot dust continuum size measured by OI is in general about 2--2.5 times larger than 
that measured by the RM time lag.  It has been attributed to the difference between 
the flux-weighted radius and response-weighted radius of the innermost hot dust 
\citep{Koshida2014}.  Moreover, \cite{SobrinoFigaredo2020} argue that the large 
\roi/\rrm\ ratio can be explained by the ``foreshortening effect'' of 
a bowl-shape dust structure \citep{PozoNunez2014,Oknyansky2015,Ramolla2018}.  
In this section, we adopt 
a simple model of hot dust emission to explore how the observed \roi/\rrm\ can 
be used to constrain the model.  Our goal is to obtain qualitative properties of 
the hot dust structure based on the samples of OI and RM observations, while 
more quantitative measurements of the hot dust structure of individual sources 
must be obtained by modeling of the OI and RM data in detail.

\begin{figure*}[htb!]
\centering
\includegraphics[width=\textwidth]{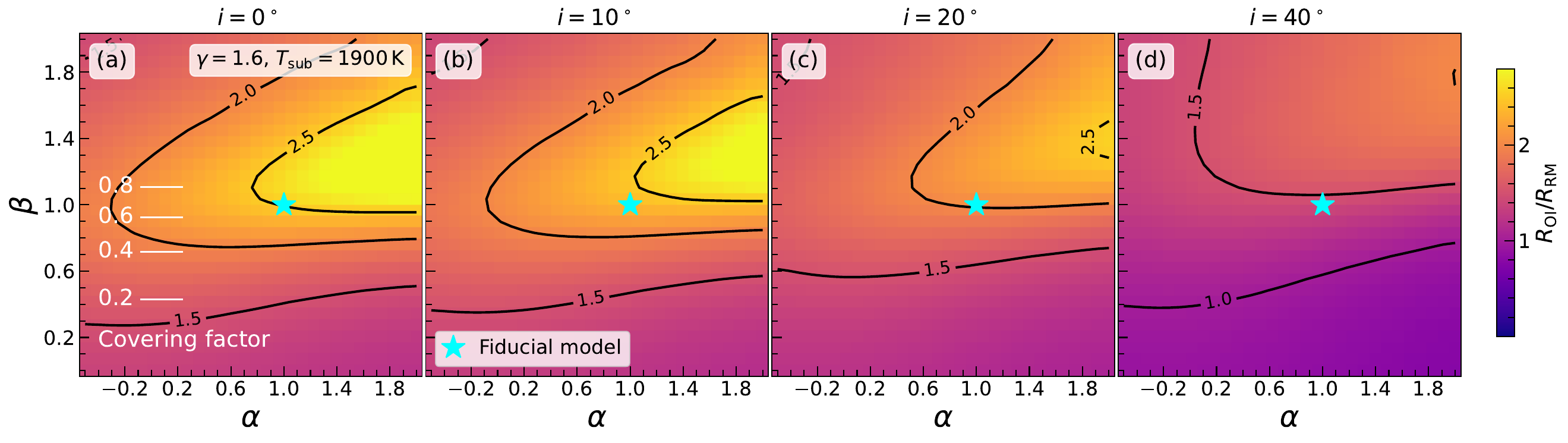}
\caption{Explore the parameter space of the hot dust structure model. The color maps show 
the derived \roi/\rrm\ ratio based on different $\alpha$ and $\beta$ in grids.  
The four columns present the model with different inclination angles ($i$ from 
0\degree\ to 40\degree).  The geometric covering factor of the model is labeled 
on the left.  A fiducial model, which can qualitatively explain our observed 
$\roi/\rrm \approx 2$, is indicated by the cyan star.}
\label{fig:parmap}
\end{figure*}

We incorporate a simple model as described in \cite{Guise2022}.  The hot dust 
emission comes from a 2-D surface which can be flaring above the midplane as 
shown in Figure~\ref{fig:sketch}.  The model is flexible enough to generate 
the bowl-shape structure which was proposed previously 
\citep[e.g.][]{Kawaguchi2010,Goad2012} and supported by RM observations 
\citep[e.g.][]{PozoNunez2014,Oknyansky2015,Ramolla2018,SobrinoFigaredo2020}.  
\textcolor{black}{
The model assumes that the dust distribution is optically thick in the vertical direction, so that the IR emission is dominated by the surface of the structure facing the observer.} 
The details of 
the model are summarized in Appendix~\ref{apd:torus}.  Briefly, the model 
consists of a large number of dust clouds, each in thermal equilibrium and 
radiating as blackbodies.  The cloud radial distribution is controlled by 
a power-law index $\alpha$, while cloud heights above the midplane are 
controlled by another power-law index $\beta$.  A higher $\alpha$ means more 
emission from the outer region, while the curved surface will become steeper 
with a larger $\beta$.  The model is observed at an inclination angle $i$.  
Only one side of the dust emission can be observed because the other side, which 
is behind the midplane, is fully obscured in NIR (See Appendix~\ref{apd:torus} 
for adopted dust temperature and spectral energy distribution (SED)).  
As illustrated by Figure~\ref{fig:sketch} (assuming $i=0\degree$ with the line 
of sight downward from the top for simplicity), the bowl-shape (one-sided) dust 
emission naturally leads to an RM time lag smaller than the projected size of 
the hot dust, the ``foreshortening effect'' (\citealt{SobrinoFigaredo2020}; and 
references therein).  To account for the observational effect that we measure 
the size of the hot dust interferometrically, we simulate the $V^2$ of 
the model assuming a typical size, distance, and baseline lengths of our targets. 
We measure the \roi\ of the simulated $V^2$ in the same way as our targets,  
by fitting a Gaussian model and converting the FWHM to a ring radius (see details in Section~\ref{sec:size}); we do not apply additional correction for the central point source, as the model does not incorporate its flux contribution. 
Meanwhile the time lag ($\rrm \equiv c\trm$) 
is calculated as the flux-weighted mean time lag of all the dust clouds.  
Both the sizes \roi\  and $\rrm $ are in units of the dust sublimation radius $r_{\mathrm{sub}}$. 
Exploration of the parameter space shows that such a model can easily 
reproduce $\roi/\rrm \approx 2$ while other properties of the model match 
various independent observational constraints.

As shown in Figure~\ref{fig:parmap}, we calculate the \roi/\rrm\ ratio of 
the model with $-0.5<\alpha<2.0$ and $0<\beta<2.0$ viewed with inclinations 
of 0\degree, 10\degree, 20\degree, and 40\degree.  To guide the eye, we 
highlight a fiducial model with $\alpha=1.0$ and $\beta=1.0$, which corresponds 
to a conical structure.  The corresponding 
\roi/\rrm\ decreases from around 2.5 to around 1.5 when the inclination 
increases from 0\degree\ to 40\degree.  At fixed inclination, the ratio 
increases when $\alpha$ and $\beta$ increase because more emission comes  
further away from the midplane and the foreshortening effect is stronger.  
When $\alpha \gtrsim 1$, the \roi/\rrm\ ratio mainly depends on $\beta$ because 
the hot dust emission mainly comes from the outer edge of the model.  

The model parameters can be further constrained by our other knowledge from 
observation and theory.  Within our preferred parameter space, \rrm \ typically falls within 1--2 times $r_{\mathrm{sub}}$, consistent with theoretical expectations and previous observations \citep[e.g.][]{Kishimoto2007, Koshida2014}. 
The spectral energy distribution 
(SED) of the hot dust emission has also been studied by many observational 
\citep[e.g.][]{Nenkova2008b,Mor2012,Lani2017,Shangguan2018,Zhuang2018} and 
theoretical works \citep[e.g.][]{Fritz2006,Nenkova2008a,Honig2010,Honig2017}.
The SED of the hot dust emission is affected by $\alpha$ and $\beta$ because 
the dust grains further from the heating source have lower temperature.  
As discussed in Appendix~\ref{apd:torus}, the NIR colors of 
our model are largely consistent with the radiation transfer model (CAT3D; 
\citealt{Honig2017}) when $-0.5<\alpha<2.0$ and $0.5<\beta<1.5$.
 Moreover, we calculate the geometric covering factor of the bowl-shape 
structure based on the maximum $h/r$ of the clouds, 
$c_f = (h/r)_\mathrm{max} / \sqrt{1+(h/r)_\mathrm{max}^2}$.  The $c_f$ only 
depends on $\beta$ because we assume the 2-D surface is fully filled for 
simplicity in Figure~\ref{fig:parmap}. 
When $\beta$ increases,
the accretion disk is more likely to be obscured by the dust, namely 
the covering factor $c_f$ is higher, because more solid angle is covered by the dusty structure.  
Various observations indicate the dust covering factor is typically 0.6-0.8, for example from the fraction of obscured AGNs \citep{Huchra1992,Ricci2017}, dust reprocessed AGN luminosity fraction \citep{Stalevski2016}, and modeling the X-ray spectrum 
\citep{Zhao2021}.  
We find that in the model, the geometric covering factor is 
$\sim 0.7$ when $\beta \approx 1.0$, consistent with the observations. 

The ratio \roi/\rrm\ in our sample appears relatively constant across the entire luminosity range, yet intriguing variations might arise depending on other properties of the AGN. 
\cite{Ricci2017} found that the dust covering factor of AGNs with a high Eddington ratio (e.g. $\ledd \gtrsim 0.1$) is much lower than the low 
Eddington ratio AGNs.  In the context of our model, this means $\beta$ decreases 
significantly when $\ledd>0.1$.  Since the \roi/\rrm\ is sensitive to $\beta$, 
we expect to find low \roi/\rrm\ for sources with high $\ledd$.
 However, the current small sample size does not allow for robust statistical analysis on this. 
 More observations of high-Eddington ratio AGNs will be valuable in exploring this dependence. 
In addition,  the model suggests that there is an inclination dependence of \roi/\rrm. It would be also interesting to investigate the relationship between inclination and \roi/\rrm\ by gathering independent estimations of inclination for a statistically significant sample.

In summary, we conducted a heuristic search of the parameter space 
of a simple hot dust model to find where \roi/\rrm\ is around 2.  We find a 
bowl-shape dust emitting structure can explain the observed values with the ratio 
primarily affected by the inclination angle and $\beta$ of the model.  Moreover, 
the covering factor of the preferred model is consistent with other 
independent observations.  Our modeling illustrates that combining the OI and RM 
observations is powerful to constrain the structure of the hot dust emission.  
The geometric distance based on the joint analysis of the OI and RM observations 
should account for variation of inclination and opening angle of the bowl-shape 
for individual sources.

\section{Summary}
\label{sec:sum}

\begin{enumerate}
\item We present new measurements of the hot dust structure sizes of 14 type 1 AGNs from VLTI/GRAVITY interferometric observation. The typical FWHM is $\sim$0.7 mas, comparable to previous near-infrared sizes, and 10-20 times smaller than mid-infrared sizes in the literature. 
\item  We compiled a sample of 25 AGNs at $z\lesssim 0.2$ with hot dust sizes measured by optical/infrared interferometry, covering four orders of magnitude in luminosity, to study the \rl\ relation.  We also compile a sample of 29 AGNs at comparable redshifts with RM measured continuum sizes for comparison.  The analysis shows a tight correlation between the size of the innermost hot dust structure and AGN luminosity, with an intrinsic scatter of less than $0.2$ dex, consistent for both OI and RM measured relations. 
\item  The slopes of the OI and RM \rl\ relations are consistent with the canonical value of 0.5 ($R\propto L^{0.5}$) within 1-$\sigma$ if we adopt more accurately derived bolometric luminosity with non-linearity correction. Meanwhile, we find the slope of the relation deviates from 0.5 most significantly with the optical luminosity.  We emphasize that {proper} bolometric corrections should be used when one investigates the \rl\ relation.
\item Converted to physical sizes, our direct measurements from GRAVITY show an offset of a factor 2 compared to the equivalent relation derived through reverberation mapping. We use a simple model to explore dust structure geometry, and conclude that a bowl-shape hot dust structure could explain the size ratio in harmony with other physical constraints. 
\end{enumerate}

\begin{acknowledgements}
This research has used the NASA/IPAC Extragalactic Database (NED), operated by the California Institute of Technology, under contract with the National Aeronautics and Space Administration. This research has used the SIMBAD database, operated at CDS, Strasbourg, France.
\end{acknowledgements}

%
%

\bibliographystyle{aa}
\bibliography{grav_agn_dust}

\begin{appendix}
\section{Bolometric luminosity correction}
\label{apd:lbol}

\begin{table*}[htb]
\caption{Properties of the AGNs used in this work.}
\label{tab:lbol_use}
\begin{center}
\resizebox{\textwidth}{!}{%
\begin{tabular}{l c c c c c  c c c c r}
\hline
\hline
\noalign{\smallskip}
Source   & $z$  
& {\lxray}  
&  {\lopt} 
& {\lmir}
& { $\log L_{\rm bol, 14-195 keV}$} 
& { $\log L_{\rm bol, 5100 \AA}$} 
& { $\log L_{\rm bol, 12 \ \mu m}$} 
& { $\log L_{\rm bol}$ } 
& { $\log M_{\rm BH}$} 
& { $\log \lambda_{\rm Edd}$}  
\\

 &  
& ($\mathrm{erg~s^{-1}}$)  
& ($\mathrm{erg~s^{-1}}$)  
& ($\mathrm{erg~s^{-1}}$) 
&  ($\mathrm{erg~s^{-1}}$) 
&  ($\mathrm{erg~s^{-1}}$) 
&  ($\mathrm{erg~s^{-1}}$) 
&  ($\mathrm{erg~s^{-1}}$) 
&  ($\mathrm{M_{\odot}}$) 
&   
\\

 (1)  & (2)  
& (3)  
& (4) 
&  (5)
&  (6)
&  (7)
&  (8)
&  (9)
&  (10)
& (11)  
\\

\noalign{\smallskip}
\hline
3C 120 & $0.033 $  & $ 44.4 $  & $ 44.0 $  & $ 44.2 $  & $ 45.3 $  & $44.9 $  & $ 44.9 $  & $45.3 $  & $7.5 $  & $-0.29 $  \\
3C 273 & $0.158 $  & ...  & $ 45.9 $  & $ 45.7 $  & ...  & $46.6 $  & $ 46.5 $  & $46.6 $  & $8.5 $  & $0.03 $  \\
Akn 120 & $0.033 $  & $ 44.3 $  & $ 43.9 $  & $ 44.2 $  & $ 45.2 $  & $44.8 $  & $ 44.9 $  & $45.2 $  & $8.4 $  & $-1.40 $  \\
ESO 323-G77 & $0.015 $  & $ 43.2 $  & $ 43.1 $  & $ 43.7 $  & $ 44.0 $  & $44.1 $  & $ 44.3 $  & $44.0 $  & $7.1 $  & $-1.20 $  \\
GQ Com & $0.165 $  & $ 44.9 $  & $ 44.6 $  & ...  & $ 45.9 $  & $45.5 $  & ...  & $45.9 $  & $8.3 $  & $-0.54 $  \\
H0507+164 & $0.018 $  & $ 43.8 $  & $ 42.6 $  & ...  & $ 44.7 $  & $43.6 $  & ...  & $44.7 $  & $7.0 $  & $-0.41 $  \\
HE 1029-1401 & $0.086 $  & $ 44.8 $  & $ 44.6 $  & ...  & $ 45.8 $  & $45.5 $  & ...  & $45.8 $  & $8.7 $  & $-1.08 $  \\
IC 4329A & $0.016 $  & $ 44.3 $  & $ 43.5 $  & $ 44.2 $  & $ 45.1 $  & $44.4 $  & $ 44.9 $  & $45.1 $  & $8.2 $  & $-1.13 $  \\
IRAS 03450+0055 & $0.032 $  & ...  & $ 43.9 $  & ...  & ...  & $44.8 $  & ...  & $44.8 $  & $7.8 $  & $-1.07 $  \\
IRAS 09149-6206 & $0.057 $  & $ 44.4 $  & $ 45.0 $  & $ 45.0 $  & $ 45.3 $  & $45.8 $  & $ 45.7 $  & $45.3 $  & $8.1 $  & $-0.83 $  \\
IRAS 13349+2438 & $0.108 $  & ...  & $ 45.0 $  & $ 45.6 $  & ...  & $45.8 $  & $ 46.3 $  & $45.8 $  & $7.8 $  & $-0.12 $  \\
MCG+08-11-011 & $0.021 $  & $ 44.1 $  & $ 43.3 $  & ...  & $ 45.0 $  & $44.3 $  & ...  & $45.0 $  & $7.7 $  & $-0.82 $  \\
MCG-6-30-15 & $0.008 $  & $ 43.0 $  & $ 41.6 $  & $ 43.1 $  & $ 43.7 $  & $42.7 $  & $ 43.7 $  & $43.7 $  & $6.6 $  & $-1.04 $  \\
Mrk 110 & $0.035 $  & $ 44.2 $  & $ 43.7 $  & ...  & $ 45.1 $  & $44.6 $  & ...  & $45.1 $  & $7.1 $  & $-0.10 $  \\
Mrk 1239 & $0.020 $  & ...  & $ 44.5 $  & $ 44.1 $  & ...  & $45.4 $  & $ 44.8 $  & $45.4 $  & $6.4 $  & $0.85 $  \\
Mrk 231 & $0.042 $  & ...  & $ 45.0 $  & $ 45.2 $  & ...  & $45.8 $  & $ 45.9 $  & $45.8 $  & $7.9 $  & $-0.16 $  \\
Mrk 335 & $0.026 $  & $ 43.5 $  & $ 43.8 $  & ...  & $ 44.3 $  & $44.7 $  & ...  & $44.3 $  & $6.9 $  & $-0.73 $  \\
Mrk 509 & $0.034 $  & $ 44.5 $  & $ 44.2 $  & $ 44.3 $  & $ 45.4 $  & $45.1 $  & $ 45.0 $  & $45.4 $  & $8.2 $  & $-0.90 $  \\
Mrk 590 & $0.026 $  & $ 43.5 $  & $ 43.5 $  & $ 43.6 $  & $ 44.2 $  & $44.4 $  & $ 44.3 $  & $44.2 $  & $7.6 $  & $-1.43 $  \\
Mrk 6 & $0.019 $  & $ 43.7 $  & $ 43.6 $  & ...  & $ 44.6 $  & $44.5 $  & ...  & $44.6 $  & $8.0 $  & $-1.56 $  \\
Mrk 744 & $0.009 $  & $ 42.5 $  & $ 41.8 $  & ...  & $ 43.2 $  & $42.9 $  & ...  & $43.2 $  & $7.4 $  & $-2.26 $  \\
Mrk 79 & $0.022 $  & $ 43.7 $  & $ 43.7 $  & ...  & $ 44.6 $  & $44.6 $  & ...  & $44.6 $  & $7.8 $  & $-1.37 $  \\
Mrk 817 & $0.031 $  & $ 43.8 $  & $ 43.7 $  & ...  & $ 44.6 $  & $44.6 $  & ...  & $44.6 $  & $8.1 $  & $-1.53 $  \\
NGC 1365 & $0.005 $  & $ 42.7 $  & $ 41.9 $  & $ 42.8 $  & $ 43.4 $  & $43.0 $  & $ 43.4 $  & $43.4 $  & $6.3 $  & $-1.00 $  \\
NGC 3227 & $0.004 $  & $ 42.6 $  & $ 42.2 $  & $ 42.2 $  & $ 43.3 $  & $43.2 $  & $ 42.8 $  & $43.3 $  & $7.1 $  & $-1.94 $  \\
NGC 3516 & $0.009 $  & $ 43.3 $  & $ 42.8 $  & ...  & $ 44.1 $  & $43.8 $  & ...  & $44.1 $  & $7.8 $  & $-1.81 $  \\
NGC 3783 & $0.010 $  & $ 43.6 $  & $ 43.0 $  & $ 43.6 $  & $ 44.4 $  & $44.0 $  & $ 44.2 $  & $44.4 $  & $7.6 $  & $-1.27 $  \\
NGC 4051 & $0.002 $  & $ 41.7 $  & $ 41.9 $  & $ 42.2 $  & $ 42.3 $  & $43.0 $  & $ 42.7 $  & $42.3 $  & $5.7 $  & $-1.50 $  \\
NGC 4151 & $0.003 $  & $ 43.1 $  & $ 42.1 $  & $ 42.9 $  & $ 43.9 $  & $43.2 $  & $ 43.5 $  & $43.9 $  & $7.7 $  & $-1.93 $  \\
NGC 4593 & $0.008 $  & $ 43.2 $  & $ 42.6 $  & $ 43.0 $  & $ 43.9 $  & $43.6 $  & $ 43.6 $  & $43.9 $  & $7.3 $  & $-1.50 $  \\
NGC 5548 & $0.017 $  & $ 43.8 $  & $ 43.3 $  & $ 43.3 $  & $ 44.6 $  & $44.3 $  & $ 44.0 $  & $44.6 $  & $8.2 $  & $-1.68 $  \\
NGC 7469 & $0.016 $  & $ 43.6 $  & $ 43.5 $  & $ 43.9 $  & $ 44.5 $  & $44.4 $  & $ 44.5 $  & $44.5 $  & $7.6 $  & $-1.26 $  \\
NGC 7603 & $0.029 $  & $ 44.0 $  & $ 44.4 $  & ...  & $ 44.9 $  & $45.3 $  & ...  & $44.9 $  & $8.4 $  & $-1.65 $  \\
PDS 456 & $0.184 $  & ...  & $ 46.3 $  & $ 45.9 $  & ...  & $47.0 $  & $ 46.6 $  & $47.0 $  & $8.7 $  & $0.22 $  \\
PG 0844+349 & $0.064 $  & ...  & $ 44.2 $  & $ 44.0 $  & ...  & $45.1 $  & $ 44.8 $  & $45.1 $  & $7.7 $  & $-0.69 $  \\
PG 0953+414 & $0.234 $  & ...  & $ 45.2 $  & ...  & ...  & $46.0 $  & ...  & $46.0 $  & $8.4 $  & $-0.55 $  \\
PG 1613+658 & $0.121 $  & $ 44.7 $  & $ 44.8 $  & ...  & $ 45.6 $  & $45.6 $  & ...  & $45.6 $  & $8.8 $  & $-1.27 $  \\
PGC 50427 & $0.024 $  & $ 43.4 $  & $ 43.1 $  & ...  & $ 44.1 $  & $44.1 $  & ...  & $44.1 $  & $7.3 $  & $-1.30 $  \\
PGC 89171 & $0.027 $  & $ 43.5 $  & $ 43.9 $  & ...  & $ 44.4 $  & $44.8 $  & ...  & $44.4 $  & $7.6 $  & $-1.36 $  \\
UGC 11763 & $0.063 $  & ...  & $ 44.3 $  & $ 44.7 $  & ...  & $45.2 $  & $ 45.4 $  & $45.2 $  & $7.3 $  & $-0.23 $  \\
UGC 545 & $0.061 $  & ...  & $ 44.5 $  & $ 45.0 $  & ...  & $45.4 $  & $ 45.7 $  & $45.4 $  & $7.0 $  & $0.28 $  \\
WPVS 48 & $0.037 $  & ...  & $ 43.6 $  & ...  & ...  & $44.5 $  & ...  & $44.5 $  & $7.0 $  & $-0.57 $  \\
Z 229-15 & $0.028 $  & ...  & $ 42.9 $  & ...  & ...  & $43.9 $  & ...  & $43.9 $  & $6.9 $  & $-1.15 $  \\

\noalign{\smallskip}
\hline
\end{tabular}}
\end{center}
\textbf{Notes.}{ (1) Source name. 
(2) Redshift from NED. 
(3) X-ray 14-195 keV 70-month \textit{Swift}-BAT survey catalogue \citep{Baumgartner2013} 
(4) optical luminosity at $5100 \rm \ \AA$ $\lambda L_{\lambda}(5100 \rm \AA)$ collected from the BASS catalogue \citep{Koss2017}. 
(5) $12~ \mathrm{\mu m}$ luminosity  $\lambda L_{\lambda}(12 \rm \ \mu m)$ from \citet{Asmus2014}. 
(6) Bolometric luminosity derived from \lxray~using Equation \ref{eq:lbol_xray}. 
(7) Bolometric luminosity derived from \lopt~using Equation \ref{eq:lbol_opt}. 
(8) Bolometric luminosity derived from \lmir~using Equations \ref{eq:softx_mir}, \ref{eq:L_xray_soft} and \ref{eq:lbol_xray}. 
(9) The adopted bolometric luminosity in this work: $L_{\rm bol, 14-195 keV}$ is preferred; when it is unavailable, 
$L_{\rm bol, 5100 \AA}$ is adopted.   
(10) Black hole mass from \citet{Gravity2023}.   
(11) Eddington ratio calculated from columns (9) and (10).
}
\end{table*}

\begin{figure*}[htb!]
\centering      
\includegraphics[width =0.8\textwidth]{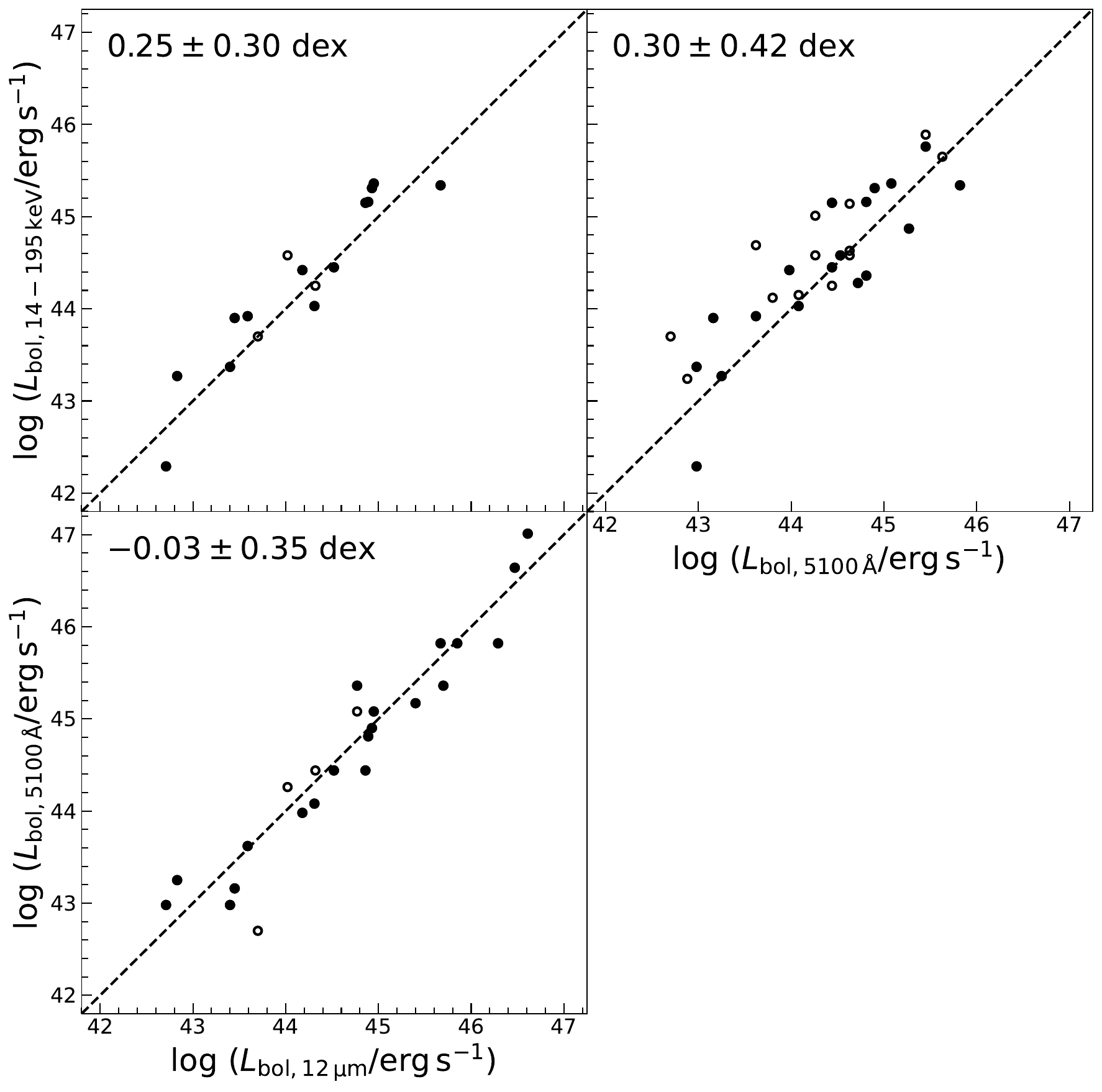}
\caption{Comparisons of \lbol \ corrections from different measurements. The solid dots highlight the sources with OI measured sizes. The dashed lines show the one-to-one relation. The median and standard deviation of $y-x$ are shown in the top-left corner in each panel. }
\label{fig:lbol_comp}
\end{figure*}



We convert each type of measured luminosity to the bolometric luminosity, using non-linear corrections, following the same method as \cite{Dexter2020}. 
We collect the 14-195 keV flux from the 70-month \textit{Swift}-BAT survey catalogue \citep{Baumgartner2013}. 
We mainly use $L_{14-195 \rm \ keV}$ to calculate the bolometric luminosity following the relation from \citet{Winter2012}, 
{which is derived using bolometric luminosities from optical-to-X-ray SED fitting \citep{Vasudevan2007, Vasudevan2009}}, 
\begin{equation}
\label{eq:lbol_xray}
\log\left(\frac{L_{\rm bol, 14-195 \ keV}}{\rm erg \ s^{-1}}\right)
= 1.1157 \log\left(\frac{L_{14-195 \rm \ keV}}{\rm erg \ s^{-1}}\right)
- 4.2280.
\end{equation}
We also use monochromatic luminosity at $5100 \rm \AA$ following the relation in \citet{Trakhtenbrot2017}, 
\begin{equation}
\label{eq:lbol_opt}
\log\left(\frac{L_{\rm bol, 5100 \ \AA}}{\rm erg \ s^{-1}}\right)
= 0.916 \log\left(\frac{\lambda L_{\lambda}(5100 \ \AA)}{\rm erg \ s^{-1}}\right)
+ 4.596.    
\end{equation}
{This relation was obtained using the bolometric correction for the $B$ band from \citet{Marconi2004} based on luminosity-dependent SED templates, and additionally assuming a constant UV-to-optical spectral slope  for the conversion from $B$ band correction to $5100 \rm \AA$. }  

For the MIR flux, we first convert $12 \rm \ \mu m $ flux $f_{12 \rm \ \mu m}$ to $2-10 \rm \ keV$ flux $f_{2-10 \rm \ keV}$ following \citet{Asmus2011}, 
\begin{equation}
\label{eq:softx_mir}
\log\left(\frac{f_{2-10 \rm \ keV}}{\rm erg \ s^{-1} \ cm^{-2}}\right) 
= 0.89 \log\left(\frac{f_{12 \rm \ \mu m}}{\rm mJy}\right) - 12.81,
\end{equation}
then $L_{14-195 \rm \  keV}$ is calculated from $2-10 \rm \ keV$ luminosity using the relation from \citet{Winter2009}, 
\begin{equation}
\label{eq:L_xray_soft}
\log\left(\frac{L_{14-195 \rm \  keV}}{\rm erg \ s^{-1}}\right)
= 0.94 \log\left(\frac{L_{2-10 \rm \ keV}}{\rm erg\ s^{-1}}\right) + 2.91.  
\end{equation}
{Both of the relations are established empirically. F}inally the MIR-based bolometric luminosity $L_{\rm bol, 12 \ \mu m}$ is obtained using Equation \ref{eq:lbol_xray}.

We list the results of the bolometric luminosities calculated from the above methods in Table \ref{tab:lbol_use}.
We compare the three different bolometric luminosities in Figure \ref{fig:lbol_comp}. We find that the differences are within $0.3$ dex, and we use this value as the uncertainty of bolometric luminosities.  The variability effect is partially captured within this uncertainty since the luminosities at different bands are collected at different times.

The bolometric luminosities adopted for studying the \rlbol~ relation in this work are listed in column (10) of Table \ref{tab:lbol_use}. 
We prioritize \lxray~-based bolometric luminosity; when it is unavailable, \lopt~-based calculation is used. 
{ This preference is due to the \lxray-based correction being established with bolometric luminosities reliably determined from SED fitting, whereas additional SED model assumptions were required for the \lopt-based calculations, potentially leading to extra uncertainties. The \lmir-based bolometric calculation incorporated two empirical relations in addition to the one between \lbol~and 14--195 keV luminosity, making it potentially less reliable; therefore, we use it only for comparative evaluations. }
For AGNs common with with those studied by \citet{Prieto2010}, the bolometric luminosities we adopted are consistent with those derived from nuclear SEDs by those authors.


%

\section{The torus model}
\label{apd:torus}

\begin{figure}[htb!]
\centering
\includegraphics[width=0.48\textwidth]{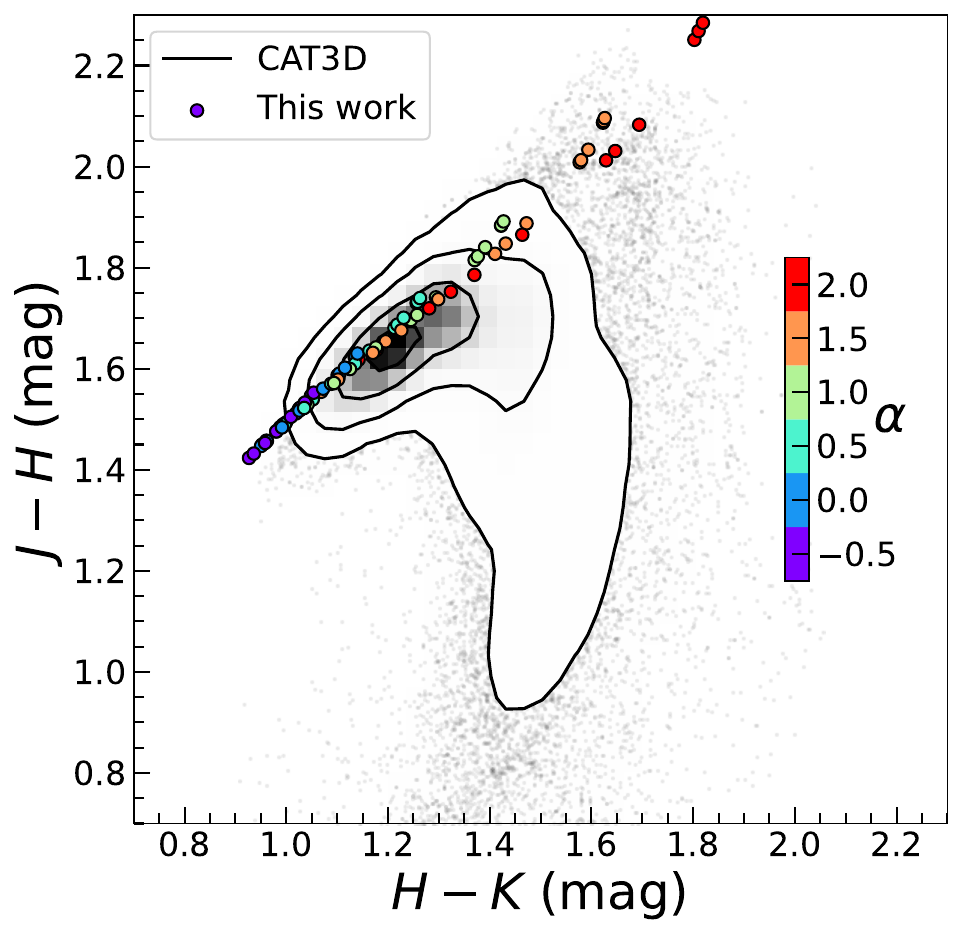}
\caption{The $J-H$ and $H-K_s$ relation of our torus model (color-coded) 
comparing that of the CAT3D-wind model (in black).  We calculated our torus 
model with $-0.5<\alpha<2.0$, $0.5<\beta<1.5$, and $i<40\degree$.  
We include the CAT3D-wind model SEDs with $i<45\degree$ for comparison.}
\label{fig:ccd}
\end{figure}

Following \cite{Guise2022}, we build a torus model to simulate the hot dust 
emission and time lag in $K$ band.  Our primary goal is to investigate whether 
the observed $\roi/\rrm \approx 2$ can be explained by a simple model.  
The model has been discussed comprehensively in \cite{Guise2022}.  We briefly 
summarize its key points and clarify our treatments that are different from 
\cite{Guise2022}. 

The torus model consists of a large number of dust clouds randomly generated to 
form a 2D surface.  The radial distribution of the clouds follows a power-law 
probability density function (PDF),
\begin{equation}\label{eq:tora}
\mathrm{PDF}(r) \propto \left(\frac{r}{r_{\rm sub}}\right)^{\alpha}, 
\end{equation}
where $r$ is radial distance of cloud from the center, \rsub\ is 
the sublimation radius of the dust, and power-law index, $\alpha$, is a primary 
free parameter of the model. 
The underlying radial density of clouds thus follows a power-law with an index of $\alpha-1$; a value of $\alpha=1$ means a flat density profile at any radius, while a higher $\alpha$ means more clouds distribute to larger distances. We adopted the maximum radius of the clouds to be 20 times that of \rsub.  However,  \roi/\rrm\ is not sensitive to the maximum radius 
because both \roi\ and \rrm\ increase with the maximum radius.  The height of 
the clouds above the midplane, $h$, follows a power-law function, 
\begin{equation}\label{eq:torb}
h = r_\mathrm{sub} \left[ \left(\frac{r} {r_{\rm sub}}\right)^\beta -1 \right], 
\end{equation}
where the power-law index $\beta$ is another primary parameter of this 
model.  The model is close to a flat disk when $\beta$ is close to 0, and is close to parabolic when $\beta=2$.  Although the dust torus may 
have a more complicated 3D structure \citep{Honig2019}, the $K$-band emission 
is expected to be emitted from the hottest dust close to the surface of 
the torus facing the radiation from the accretion disk.  Considering the effect 
of illumination, we include the emission weight of the clouds,
\begin{equation}\label{eq:torc}
\kappa = 0.5 (1-\cos{\psi}), 
\end{equation}
where $\psi$ is the angle between observer's line of sight and cloud's line of 
sight to the center from the origin (the radiation source).  Each dust cloud is 
assumed to have black body emission in a equilibrium state according to 
the absorbed emission from the central radiation source, so the dust temperature 
is,
\begin{equation}\label{eq:tord}
T(r) = T_{\rm sub} \left(\frac{r}{r_{\rm sub}}\right)^{\frac{-2}{4+\gamma}},
\end{equation}
where $r$ is the radius of the dust cloud, \tsub\ is the sublimation temperature, 
and $\gamma$ is the dust IR opacity power-law index which is around 1--2 for 
interstellar dust.  We adopt $\gamma=1.6$ for typical astronomical dust 
following \cite{Barvainis1987}, while we find that our conclusions are not 
sensitive to the adopted $\gamma$.  We choose to use $\tsub=1900$~K, instead of 
1500~K adopted by \cite{Barvainis1987}, because recent observations found 
increasing evidence of a higher sublimation temperature due to the graphite dust 
grains \citep{Mor2012,Honig2017}.  Our model $\roi/\rrm$ is not very sensitive 
to the \tsub, but as discussed later, we find our model provides consistent 
$JHK$ color when adopting $\tsub=1900$~K.

In order to calculate the \roi, we first simulate the observed visibility of 
the dust torus model with the realistic baseline lengths of GRAVITY and apply 
the same fitting method as described in Section~\ref{sec:size}.  The torus time 
lag is calculated as the flux-weighted mean time lag of each 
cloud.\footnote{The time lag of the $i$th cloud is 
$\tau_i = r_i (1 - \cos\psi_i)$.}  In this work, we use the model in a heuristic 
manner to investigate whether there is a parameter range that can explain our 
observed \roi/\rrm.  We explored mainly the parameter space defined in 
\cite{Guise2022}.  We found that $\alpha \approx 1.0$ is preferred to 
obtain $\roi/\rrm \approx 2$, much larger than the parameter range defined in 
\cite{Guise2022} ($-5.5<\alpha<-0.5$).  The $\alpha$ controls the radial 
distribution of the dust clouds so it influences the SED of the model.  
Our preferred $\alpha$ ranges are not necessary the same as theirs since we are focusing 
on $K$ band observations, while \cite{Guise2022} are working in longer 
wavelengths.  We compare the $JHK$ colors predicted by our model to the CAT3D 
model \citep{Honig2017}, one of the state-of-the-art torus models considering 
different temperatures of the graphite and silicon dust and the polar wind 
structure.  We focus on the $JHK$ colors because, unlike CAT3D, our model only 
consider the hottest dust in the torus surface.  As shown in 
Figure~\ref{fig:ccd}, the $J-H$ and $H-K$ colors of our simple model matches 
those of the CAT3D model\footnote{Some CAT3D models show a low $J-H$ color 
(e.g. at $1.2 < H-K < 1.8$) primarily because the accretion disk emission 
contaminates the color when the torus emission is very red.} in the parameter 
ranges that we adopt in this work, in particular, $-0.5<\alpha<2$.  We also find 
that our model will become much redder than the CAT3D model if we use 
$\tsub=1500$~K, likely because that the CAT3D model includes the graphite dust 
with the sublimation temperature at 1900~K.  The comparison with $JHK$ colors 
suggest that the dust distribution of our model is reasonable.  The conclusion 
is not sensitive to the choice of the radiative transfer model as long as 
the \tsub\ is assumed consistently.  We prefer to compare our model colors with 
the radiative transfer models over the real observation because the observed AGN 
SED are contaminated by the host galaxy which is usually bright in NIR.  In summary, we confirm that our adopted parameter ranges  align with the theoretically expected color of the torus.

\end{appendix}

\end{document}